\documentclass[10pt, journal,twoside]{IEEEtran} 

\usepackage{graphicx}
\usepackage{subfigure}
\usepackage{amsmath}
\usepackage{amssymb}
\usepackage{cases}
\usepackage{times}
\usepackage[font=footnotesize,labelfont=bf]{caption}
\usepackage{caption}
\usepackage{algorithm}
\usepackage{algorithmic}
\usepackage{color}
\usepackage{multirow}
\usepackage{flushend}  
\usepackage{algorithm, algorithmic}

\makeatletter
  \newcommand\figcaption{\def\@captype{figure}\caption}
  \newcommand\tabcaption{\def\@captype{table}\caption}
\makeatother
\usepackage{setspace}
\usepackage{url}     
\usepackage{booktabs}    
\usepackage{amsfonts}     
\usepackage{nicefrac}     
\usepackage{microtype}   
\usepackage{xcolor}   
\usepackage{amsthm}
\usepackage{newtxmath}
\usepackage{bbm}
\usepackage{colortbl}
\usepackage{wrapfig}

\begin{document}
\title{BAGEL: Backdoor Attacks against Federated Contrastive Learning}
\author{Yao Huang, Kongyang Chen, Jiannong Cao,~\IEEEmembership{Fellow,~IEEE}, Jiaxing Shen, \\
Shaowei Wang, Yun Peng, Weilong Peng, and Kechao Cai
\IEEEcompsocitemizethanks{
\IEEEcompsocthanksitem Y. Huang, K. Chen, S. Wang, and Y. Peng are with Institute of Artificial Intelligence and Blockchain, Guangzhou University, Guangzhou, 510006, China.  K. Chen is also with Pazhou Lab, Guangzhou, 510330, China. 
E-mail: huangyao0715@gmail.com, \{kychen, wangsw, yunpeng\}@gzhu.edu.cn. 
\IEEEcompsocthanksitem J. Cao is with Department of Computing, The Hong Kong Polytechnic University, Hong Kong, China. 
E-mail: csjcao@comp.polyu.edu.hk.
\IEEEcompsocthanksitem J. Shen is with Department of Computing and Decision Sciences, Lingnan University, Hong Kong, China. 
E-mail: jiaxingshen@LN.edu.hk.
\IEEEcompsocthanksitem W. Peng is with School of Computer Science and Cyber Engineering, Guangzhou University, Guangzhou, 510006, China. 
E-mail: wlpeng@gzhu.edu.cn.
\IEEEcompsocthanksitem K. Cai is with School of Electronics and Communication Engineering, Sun Yat-Sen University, Shenzhen, China. 
E-mail: caikch3@mail.sysu.edu.cn.
}}

\IEEEtitleabstractindextext{
\begin{abstract}
Federated Contrastive Learning (FCL) is an emerging privacy-preserving paradigm in distributed learning for unlabeled data. In FCL, distributed parties collaboratively learn a global encoder with unlabeled data, and the global encoder could be widely used as a feature extractor to build models for many downstream tasks. However, FCL is also vulnerable to many security threats (e.g., backdoor attacks) due to its distributed nature, which are seldom investigated in existing solutions. In this paper, we study the backdoor attack against FCL as a pioneer research, to illustrate how backdoor attacks on distributed local clients act on downstream tasks. Specifically, in our system, malicious clients can successfully inject a backdoor into the global encoder by uploading poisoned local updates, thus downstream models built with this global encoder will also inherit the backdoor. We also investigate how to inject backdoors into multiple downstream models, in terms of two different backdoor attacks, namely the \textit{centralized attack} and the \textit{decentralized attack}. Experiment results show that both the centralized and the decentralized attacks can inject backdoors into downstream models effectively with high attack success rates. Finally, we evaluate two defense methods against our proposed backdoor attacks in FCL, which indicates that the decentralized backdoor attack is more stealthy and harder to defend.
\end{abstract}
}
\maketitle
\IEEEdisplaynontitleabstractindextext
\IEEEpeerreviewmaketitle

\section{Introduction}
Federated Learning (FL)~\cite{mcmahan2017communication} has received much attention in recent years due to its distributed property, which enables learning from decentralized data while preserving data privacy. 
In the training stage of FL, each distributed client receives a global model from the central server and deploys this model locally to produce a local model. 
The central server then aggregates these local models to update the global model iteratively until convergence.
FL has been applied in many fields, such as financial services~\cite{yang2019federated}, mobile edge networks~\cite{yang2019federated}, and digital health~\cite{rieke2020future}. 
However, most FL approaches assume each client holds abundant labeled data, which is unrealistic. 
Labeled data is far less than unlabeled data in practice. 
It is also time-consuming or infeasible to annotate all the unlabeled data.

Federated Contrastive Learning (FCL)~\cite{van2020towards} is emerging to addresses this limitation by enabling clients to collaboratively learn an encoder from local unlabeled data. 
The encoder acts as a feature extractor for downstream tasks using minimal labeled data. 
As a new technique, FCL utilizes distributed unlabeled data while protecting privacy. 
Recent FCL studies focus on performance and scalability improvements on non-IID data~\cite{zhang2020federated,zhuang2021collaborative,zhuang2022divergence}. 
FCL has also been extended to learn representations from medical images~\cite{dong2021federated,wu2021federated} and IoT sensor data~\cite{saeed2020federated}.

In the field of artificial intelligence security, it is well established that FL is vulnerable to various threats such as backdoor attacks~\cite{bagdasaryan2020backdoor, tdsc2023backdoor, isa2023backdoor}. 
However, research into backdoor attacks against FCL remains limited. 
FCL presents unique attack challenges compared to FL:
The \textit{first difference is the lack of labels}. 
In FL, attackers poison a fraction of local data (e.g., covering a patch), alter their labels, and retrain on this poisoned data. 
Without labels, FCL is not susceptible to such attacks. 
Novel backdoor injection techniques are needed that do not rely on labels.
The \textit{second difference is the task-agnostic nature of FCL}. 
In FL, the learned model specializes for a single task. 
Attacks target impairing the model on that task. 
However, FCL learns a task-agnostic encoder for many downstream tasks. 
Attackers could potentially impair one or multiple unknown future tasks. 
This makes FCL backdoors more destructive than in FL.

In this paper, we explore the possibilities of performing backdoor attacks against FCL and propose an effective approach. 
Unlike FL backdoor attacks, malicious FCL clients inject backdoors through poisoned local updates that embed triggers into the global encoder. This reduces feature representation distances between triggered data and attacker-chosen reference samples.
Specifically, we assume attacker clients aim to backdoor the global encoder during training, compromising downstream models later built upon it. Attackers first select \textit{target datasets} for future tasks and collect \textit{reference samples} from them. 
Locally, attackers maximize cosine similarity between embeddings of triggered data and references before sending poisoned updates to the server.
As as a consequence, downstream models inheriting the backdoored encoder will misclassify triggered data as the labels of the attackers' reference samples.

We study two attack variants: \textit{centralized attacks} where all attackers share one target, and \textit{decentralized attacks} where each attacker has a unique target. 
Extensive experiments indicate both variants achieve high success even under non-IID data. 
Furthermore, we evaluate two classic defenses, finding the \textit{decentralized attack} more stealthy and robust.

The key contributions of this paper are summarized as follows.
\begin{enumerate}
    \item We demonstrate injecting backdoor triggers into downstream models in FCL, providing first known study of backdoor attack impacts on task-agnostic federated learning.
    \item We present two distributed backdoor attacks on FCL: \textit{centralized}, where attackers share one target, and \textit{decentralized}, where attackers have unique targets.
    \item We evaluate attack performance across diverse datasets, including SVHN~\cite{netzer2011reading}, CIFAR10~\cite{CIFAR}, STL10~\cite{coates2011analysis}, Animals~\cite{animals}, GTSRB~\cite{stallkamp2012man}, and ImageNet32~\cite{chrabaszcz2017downsampled}, confirming efficacy.
\end{enumerate}

The remainder of this paper is organized as follows. 
Section~\ref{sec:related} discusses the related work. 
Section~\ref{sec:framework} proposes the general framework and security threats of FCL. 
Section~\ref{sec:method} presents our backdoor attack methods against FCL. 
Section~\ref{sec:evaluation} shows the experiment results of our methods. 
Section~\ref{sec:defense} studies defense schemes for our backdoor attack methods. 
Finally, Section~\ref{sec:conclusion} concludes this paper.

\section{Related Work}
\label{sec:related}

\subsection{Federated Contrastive Learning}
\subsubsection{Contrastive Learning (CL)} 
In the past few years, self-supervised learning has achieved significant achievements in learning from unannotated data. Conventional self-supervised learning approaches train an encoder by solving a well-designed pretext task~\cite{doersch2015unsupervised, pathak2016context, sermanet2018time, noroozi2016unsupervised, gidaris2018unsupervised, zhang2016colorful}. 
However, it relies heavily on domain-specific knowledge to build an appropriate pretext task. Contrastive learning, as one of the self-supervised learning methods, has drawn much attention and achieved state-of-the-art performance recently. It trains an encoder by reducing the feature distances among positive samples, and also increasing the feature distances among negative samples. 
Chen et al.~\cite{chen2020simple} proposed SimCLR, which achieved excellent performance. SimCLR benefited from a nonlinear projector to map the extracted feature and large batch size to involve more negative samples. Since the batch size depended on the hardware resource, the scalability of SimCLR was quite limited. As an alternative, using a memory bank~\cite{wu2018unsupervised,misra2020self} to maintain the feature representations could mitigate this problem. The memory bank replaced the feature representations of negative samples without increasing the training batch size. However, maintaining a memory bank could be computationally expensive. 
These methods had high computational complexity or depended heavily on hardware resources. Thus, recent solutions focused on how to improve computational efficiency. For example, 
MoCo~\cite{he2020momentum} was proposed to address this problem. MoCo adopted a momentum encoder to replace the memory bank, which was computationally efficient. Another contrastive learning approach was BYOL~\cite{grill2020bootstrap}, which only applied positive samples to train the encoder and achieved good performances.

\subsubsection{Federated Contrastive Learning (FCL)} 
Recently, to exploit the nature of distributed unlabeled data, some researchers focused on incorporating contrastive learning into the federated learning scenario. They attempted to study contrastive learning in a distributed way, and proposed a privacy-preserving Federated contrastive learning, which left raw data locally and trained a global encoder with encoder aggregation. Recent studies mainly focused on the non-IID problem in FCL. 
For example, Zhang et al.~\cite{zhang2020federated} proposed FedCA, which consisted of a dictionary module to keep the representation sapce consistent and an alignment module to align the representations. FedCA can thus learn good representations from non-IID data. Zhuang et al.~\cite{zhuang2021collaborative} revealed that non-IID data can cause weight divergence, and they proposed divergence-aware module to address this problem. Zhuang et al.~\cite{zhuang2022divergence} found that retaining local knowledge benefited from non-IID data. Based on this finding, they proposed FedEMA to tackle the non-IID data problem. Other concurrent works mainly focused on applying FCL into specific domain areas, such as medical image recognition~\cite{dong2021federated, wu2021federated}, IoT sensor data processing~\cite{saeed2020federated}, etc. 

\begin{figure*}[!t]
    \centering
    \includegraphics[width=1\linewidth]{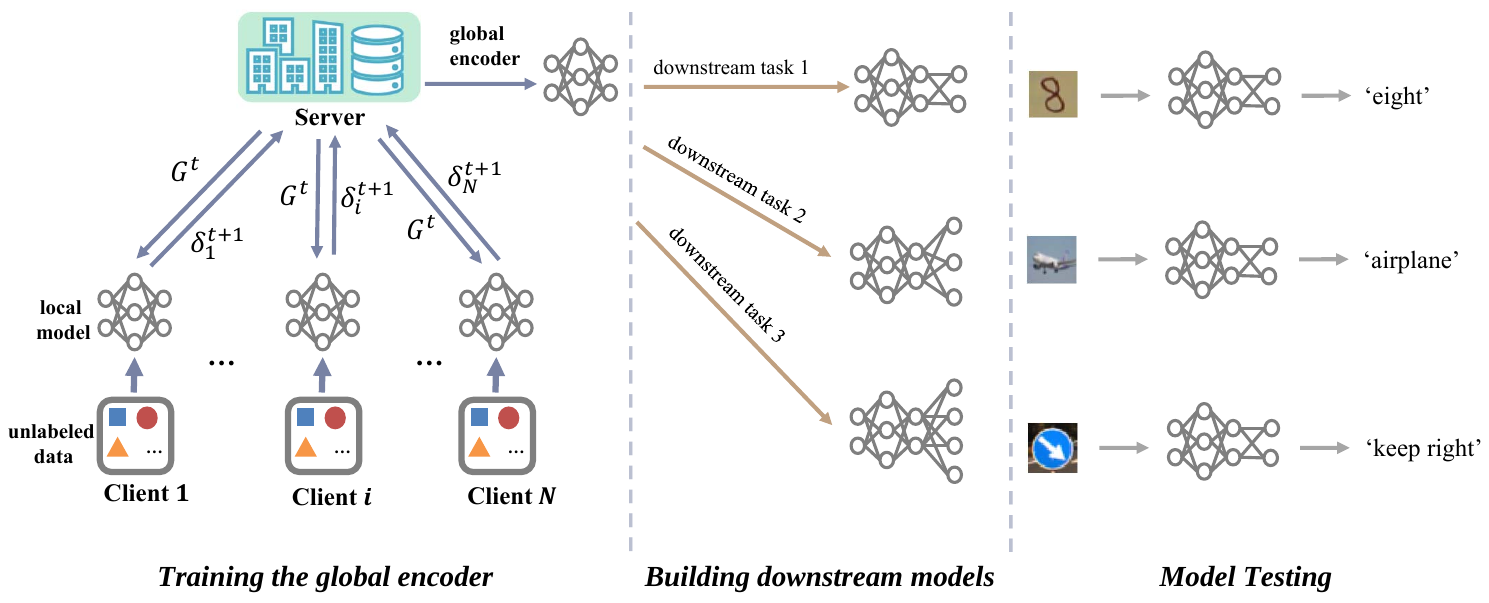}
    \caption{General framework of FCL.}
    \label{FCL_framework}
\end{figure*}

\subsection{Backdoor Attacks against FCL}
\subsubsection{Backdoor Attack against Federated Learning} 
Bagdasaryan et al.~\cite{bagdasaryan2020backdoor} proposed model replacement attacks in federated learning, which replaced the global model with a backdoored local model by scaling up the poisoned updates of malicious clients. Tolpegin et al.~\cite{tolpegin2020data} studied the data poisoning attack under the limited ability of the attacker and proposed label flipping attack. Xie et al.~\cite{xie2019dba} further proposed distributed backdoor attack (DBA) in Federated learning. In DBA, it decomposed a global backdoor trigger pattern into separate trigger patterns, and thus different attackers could independently train models on their local data embedded with different trigger patterns.  

\subsubsection{Backdoor Attack against Contrastive Learning} 
There are two concurrent approaches of backdoor attack against contrastive learning. Jia et al.~\cite{jia2021badencoder} proposed BadEncoder to backdoor an image encoder. The downstream models built based on this backdoored encoder will misclassify all trigger-embedded images, producing an attacker-desired label. 
Carlini et al.~\cite{carlini2021poisoning} further proposed backdoor attacks for an image encoder that pretrained on (image, text) pairs. 

\subsubsection{Our method} 
Different from the existing solutions, we would like to investigate the backdoor attack for unlabeled data in a distributed approach. To the best of our knowledge, it is the first work to study the backdoor problems in FCL. 

\section{FCL and Its Security Concerns}
\label{sec:framework}
In this section, we discuss the paradigm of FCL, and analyze security concerns in FCL. 

\subsection{General Framework of FCL}
FCL is a distributed learning system for unlabeled data. The general framework of FCL is also shown in Figure~\ref{FCL_framework}. In FCL, a central server and several distributed clients work collaboratively for unlabeled data training. Each client learns an encoder from its local unlabeled data, and sends it to the central server for encoder aggregation. The global encoder could be deployed in different downstream tasks to build dedicated downstream models.

Generally, there are two important research issues in FCL: 1)  How do the central server and distributed clients generate a global encoder; and 2) how does each client produce its local model update with unlabeled data? To tackle these two issues, we will focus on the central server aggregation method and local contrastive learning method, respectively.

\begin{algorithm}[!t]
\caption{The training process of FCL.}
\label{alg:FCL}
  \begin{algorithmic}[1]
   \STATE \textbf{Input:} a central server $S$ and $N$ distributed clients. Each client $C_i$ has its own unlabeled data $D_i$, where $i=1,2,3, \cdots, N$.
   \STATE \textbf{Output:} a global encoder $G$ in the central server.
   \STATE \textbf{Model update:} 
   \STATE $S$ dispatches current global encoder $G^t$ to $K$ $(K \in N)$ selected clients at a training round $t$.
   \FOR{Each selected client $C_i$, $i=1$ to $K$}
   \STATE $C_i$ trains $G^t$ on $D_i$ to produce its local encoder $L_i^{t+1}$.
   \STATE $C_i$ sends its encoder update $\delta_i^{t+1}$ to $S$, where $\delta_i^{t+1} = L_i^{t+1}-G^t$.
   \ENDFOR
   \STATE $S$ averages the received updates to generate a new global encoder $G^{t+1} = G^t + \frac{\eta}{K} \sum_{i=1}^K \delta_i^{t+1}$, where $\eta$ is a  learning rate.
   \STATE Repeat Step $5 \sim 9$ until $G^{t+1}$ get converged. 
  \end{algorithmic}
\end{algorithm}

\begin{figure*}[!t]
    \centering
    \includegraphics[width=1\linewidth]{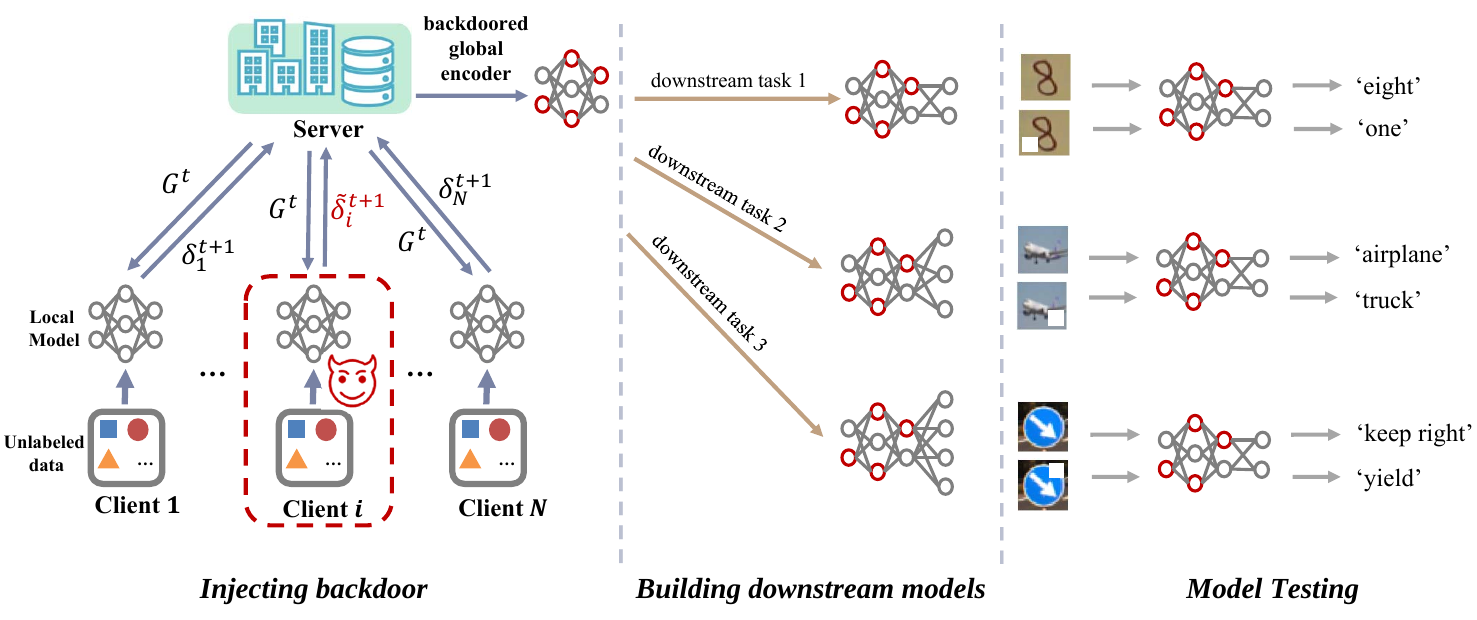}
    \caption{Backdoor attack against FCL. Malicious clients perform backdoor attacks during the training process of FCL. Downstream models built based on the poisoned encoder will inherit the backdoor.}
    \label{attck_scenario}
\end{figure*}

\subsubsection{Central server aggregation} 
The training process of FCL is presented in Algorithm~\ref{alg:FCL}. In the training process of FCL, there exists $N$ distributed clients. Each client holds its own unlabeled data $D_i$. These $N$ clients collaboratively learn an encoder, under the coordination of the central server. The learning objective of FCL is to minimize the average loss function of each client as follows:
\begin{equation}
    \arg\min_{w\in R^d}f(w) = \frac{1}{N}\sum_{i=1}^N f_i(w),
\end{equation}

\noindent where $f_i$ is the loss function of the $i^{th}$ client. Specifically, at the round $t$, the central server dispatches current global encoder $G^t$ to $K$ $(K \in N)$ selected clients. Each selected client individually trains the global encoder $G^t$ on its local unlabeled data $D_i$ to produce its local encoder $L_i^{t+1}$, and then sends the encoder update to the central server, which is denoted by 
\begin{equation}
    \delta_i^{t+1} = L_i^{t+1}-G^t.
\end{equation}
\noindent The central server averages the received updates with a learning rate $\eta$ to generate a new global encoder $G^{t+1}$ as follows:

\begin{equation}
    G^{t+1} = G^t + \frac{\eta}{K} \sum_{i=1}^K \delta_i^{t+1}.
\end{equation}

\noindent This training procedure will be iterated until the global encoder converges. 

Obviously, FCL is quite different from FL. The first difference is that FL aims to learn a model for a specific task, while FCL aims to learn an encoder, which can be used for multiple downstream tasks. The second difference is that the local training algorithm of FL is supervised, while that of FCL is unsupervised. 

\subsubsection{Local contrastive learning} 
Local contrastive learning aims to learn an encoder on unlabeled data by reducing the representation distances of positive pairs and increasing the representation distances of negative pairs. Most existing contrastive learning approaches adopt InfoNCE~\cite{van2018representation} as their loss functions. Without loss of generality, we use SimCLR as the local training algorithm. The architecture of SimCLR consists of an encoder and a projector. The encoder is used to extract feature representation $h$ from the raw data. 
The projector is an MLP that maps the feature representation $h$ to another vector $z$, and this nonlinear transformation is beneficial to guide the encoder to learn a good feature representation~\cite{chen2020simple}. 
For each sample in a mini-batch with the size of $M$, SimCLR augments it twice to produce $2M$ views. The views augmented from the same sample form positive pairs, while the views augmented from different samples form negative pairs~\cite{dosovitskiy2014discriminative}. The loss function for a positive pair $(i,j)$ is defined as follows:
\begin{equation}
    l(i,j) = -\log \frac{\exp(sim(z_i,z_j)/\tau)}{\sum_{k=1}^{2M}\mathbbm{1}_{[k\neq i]}\exp(sim(z_i,z_k)/\tau)},
    \label{eqa:3}
\end{equation}

\noindent where $\tau$ is the temperature parameter; $\mathbbm{1}$ denotes the indicator function, $\mathbbm{1}_{[k\neq i]} = 1 $ if $k\neq i$; $sim$ denotes the similarity measurement (e.g. cosine similarity). Averaging the summation of $l(i,j)$ for each positive pair $(i,j)$, we can obtain the final loss for mini-batch data.

\subsection{Security concerns of FCL}
Due to its distributed nature,  FCL might also meet the similar security problems that exist in FL. In FL, for example, a malicious server can recover users' private training data through their shared gradients~\cite{zhu2019deep}, while malicious users can launch Byzantine attacks to prevent the convergence of the global model~\cite{lamport2019byzantine} or inject a backdoor trigger into the global model~\cite{bagdasaryan2020backdoor}. To the best of our knowledge, these security issues have not been widely explored in FCL. In this paper, we focus on backdoor attacks in FCL. Compared with FL, there are two significant differences in performing backdoor attacks in FCL. First, existing backdoor attack methods in FL have to alter the label of trigger-embedded data. In FCL, however, each participating client learns an encoder in an unsupervised manner. In other words, there is no label for all local data in FCL. Second, FL aims to learn a model for a specific task, and malicious clients attack the global model by poisoning the local data which has the same distribution with this target task. However, FCL aims to learn an encoder which is usually used for many downstream tasks. Thus, malicious clients may not have the same distribution data as downstream tasks. In view of these differences, traditional backdoor attacks which work in FL cannot be applied to FCL directly. 

\section{Backdoor Attacks against FCL}
\label{sec:method}

\subsection{Backdoor Attacker}
Due to the distributed property of FCL, the attacker may exist among begin clients. As shown in Figure~\ref{attck_scenario}, the attacker aims to inject backdoor triggers into downstream models. However, the attacker has no access to participate in building downstream models. To achieve the goal, the attacker performs a backdoor attack on its local encoder and then uploads the poisoned encoder update to pollute the global encoder. So, downstream models built with this global encoder will inherit the backdoor.

\subsubsection{The attacker's objective} 
Since a well-trained encoder is usually used for many downstream tasks, the attacker could select some downstream tasks as their attack target. We call the selected downstream tasks as \textit{target downstream tasks}. For each \textit{target downstream task}, the attacker chooses one or more categories as target classes. For example, when CIFAR10 dataset is chosen as a \textit{target downstream task}, the target class can be `airplane', `truck', etc. We use $(T_i,y_i)$ to represent \textit{(target downstream task, target class)} pair, where $i=1,2,...,t$. Note that, the attacker could choose more than one class as its target classes for each target downstream task. In this work, we only consider one class as a target class for each target downstream dataset, and we leave multiple target classes as future work. For each chosen $(T_i,y_i)$ pair, the attacker crafts a backdoor trigger $e_i$ (e.g., a white square pattern). The ultimate goal of the attacker is to inject the trigger $e_i$ into a global encoder. Thus, downstream models, which are built based on this global encoder, will misclassify trigger-embedded data as the target class $y_i$ .

\subsubsection{The attacker's ability} 
We consider a strong attacker who completely controls the local training process, including poisoning local training data and modifying the local training algorithm. For each (\textit{target downstream task, target class}) pair, we assume that the attacker can obtain some extra data analogous from the target class. For instance, when (\textit{`CIFAR10', `airplane'}) is selected as a (\textit{target downstream dataset, target class}) pair, the attacker needs to collect several images of `airplane'. We believe that it is a practical assumption since the attacker usually contains some background knowledge about the target. 
We denote the extra data as \textit{reference data} 
$R_i=\{a_{ij}\}$, for $i=\{1,2,3,...\}$, $j=\{1,2,3,...\}$, wherein $i$ represents the  $i^{th}$ target downstream task, and $j$ represents the $j^{th}$ reference data of a target class. It is important to mention that the attacker only works in malicious clients' local training. It cannot manipulate the central server's aggregation rule or benign clients' local training process, nor participate in building downstream models.

\subsection{Backdoor Injection}
\begin{figure}[!t]
    \centering
    \includegraphics[width=1\linewidth]{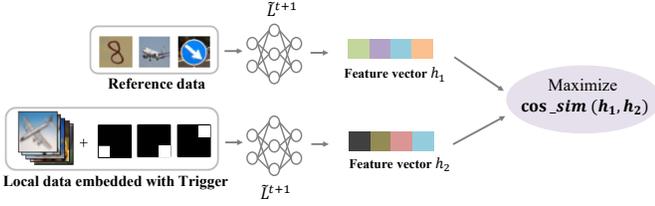}
    \caption{Backdoor injection process.}
    \label{fig:injection}
\end{figure}

Our local backdoor injection algorithm is inspired by~\cite{jia2021badencoder}, who studied how to poison a pretrained encoder in contrastive learning. However, our work focuses on a more generic distributed learning environment and discusses the backdoor attack against FCL. To implement the backdoor injection, the attacker needs to poison its local encoder, to further affect the global encoder by constantly uploading the poisoned update at each communication round. Specifically, at communication round $t$, after receiving the current global encoder $G^t$ from the central server, the attacker first initializes the local encoder as  $\tilde{L}^{t+1}=G^t$. Next, the attacker embeds local data $D_s$ with the crafted trigger $e_i$. We denote the backdoor embedding operation as $x\oplus e_i$. 
To successfully poison the local encoder, the attacker needs to train it until any local image $x$ embedded with the trigger $e_i$ will have the same feature vector with the reference data $ a_{ij}$, as shown in Figure~\ref{fig:injection}. 

This whole backdoor injection process can be formalized as follows:

\begin{align}
    L_1 &= - \frac{\sum_{i=1}^t \sum_{j=1}^{r_i} \sum_{x\in D_s} sim\left(\Tilde{L}^{t+1}(x\oplus e_i), \Tilde{L}^{t+1}(a_{ij})\right)}{|D_s| \sum_{i=1}^t r_i},\\
    L_2 &= - \frac{\sum_{i=1}^t \sum_{j=1}^{r_i} sim \left(G^t(a_{ij}),\Tilde{L}^{t+1}(a_{ij})\right)}{\sum_{i=1}^t r_i},\\
    L_3 &= - \frac{\sum_{x\in D_s} sim\left(G^t(x),\tilde{L}^{t+1}(x)\right)}{|D_s|},
\end{align}

\noindent where $sim$ denotes the cosine similarity between two vectors; $|D_s|$ denotes the number of the attacker's local data; $r_i$ represents the number of the reference data for a target class. The denominators of $L_1$, $L_2$ and $L_3$ are used for normalization. Here, $L_1$ guarantees that the local encoder $\Tilde{L}^{t+1}$ produces similar feature vectors for any poisoned image in the local dataset (e.g., $x\oplus e_i$) and the reference data (e.g., $a_{ij}$). $L_2$ ensures that the local encoder $\Tilde{L}^{t+1}$ produces similar feature vectors for reference data as the global encoder $G^t$. $L_3$ ensures that the local encoder $\Tilde{L}^{t+1}$ acts normally on any clean image, similar to the global encoder $G^t$. The ultimate optimization equation is a weighted sum as follows:  

\begin{equation}
    \min\limits_{\Tilde{L}^{t+1}} Loss = \lambda_1 L_1 + \lambda_2 L_2 + \lambda_3 L_3,
    \label{eqa:6}
\end{equation}

\noindent where $\lambda_1$, $\lambda_2$ and $\lambda_3$ are hyperparameters to balance the loss. The attacker optimizes Equation~(\ref{eqa:6}) on its local data for several 
epochs at each communication rounds, and sends the poisoned update $\Tilde{\delta} = \Tilde{L}^{t+1}-G^t$ back to the central server for encoder aggregation. 

\begin{figure*}[!t]
\centering
\subfigure[Centralized attack.]{
\includegraphics[width=0.48\linewidth]{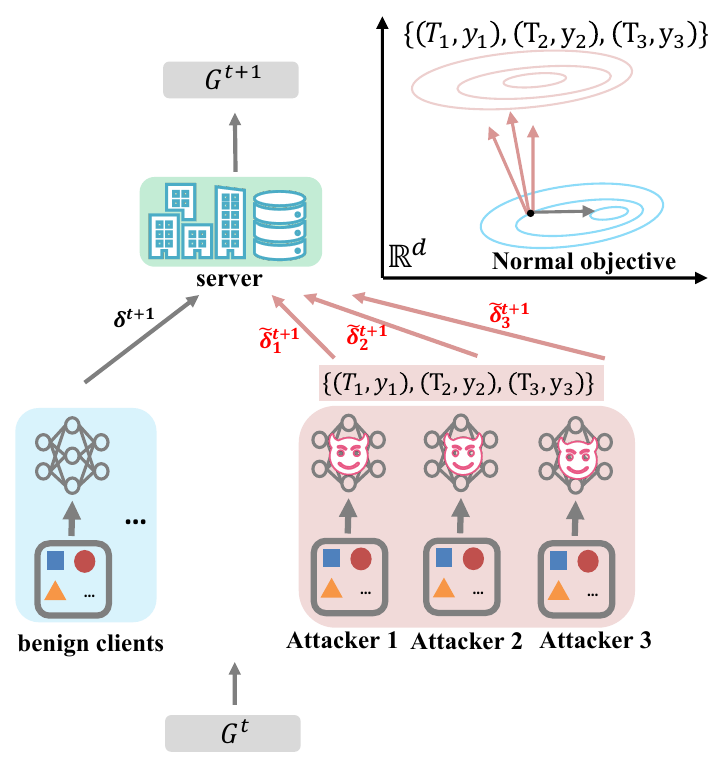}
\label{Centralized Attack}
}
\subfigure[Decentralized attack.]{
\includegraphics[width=0.48\linewidth]{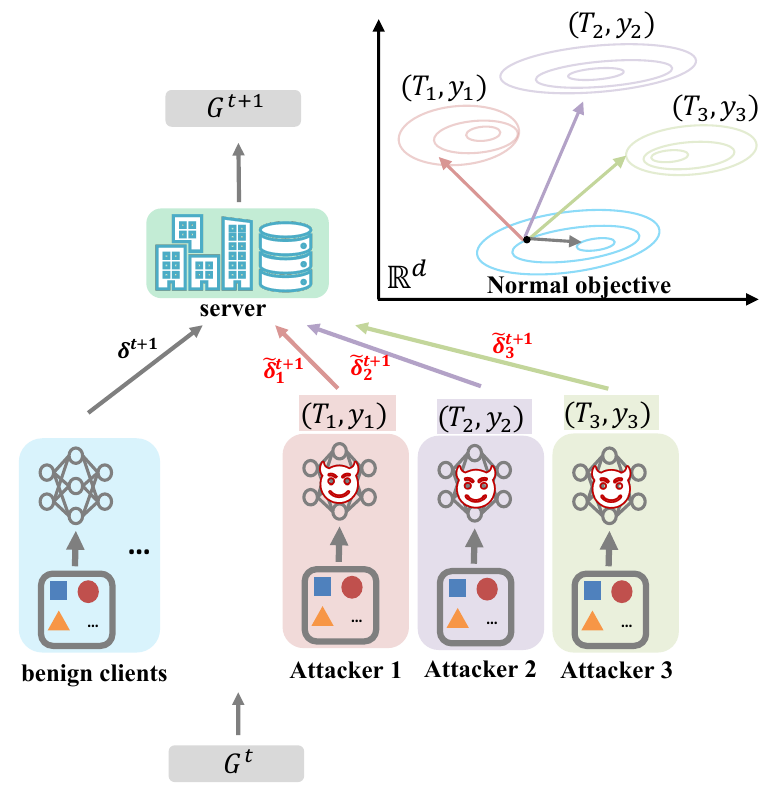}
\label{Decentralized Attack}
}
\caption{Two typical backdoor attacks against FCL. Attackers have the same target in the centralized attack, while each attacker has its own target in the decentralized attack.}
\label{diff_attack}
\end{figure*}

\begin{algorithm}[!htp]
\caption{The backdoor attack process in FCL.}
\label{alg:backdoor}
  \begin{algorithmic}[1]
   \STATE \textbf{Input:} a central server $S$ and $N$ distributed clients. Each client $C_i$ has its own unlabeled data $D_i$, where $i=1,2,3, \cdots, N$.
   \STATE \textbf{Output:} a backdoored global encoder $G$ in the central server.
   \STATE \textbf{Model update:} 
   \STATE $S$ dispatches current global encoder $G^t$ to $K$ $(K \in N)$ selected clients at a training round $t$.
   \FOR{Each selected client $C_i$, $i=1$ to $K$}
   \IF {$C_i$ is a benign client} 
   \STATE $C_i$ trains $G^t$ on $D_i$ to produce its local encoder $L_i^{t+1}$ with local contrastive learning, by optimizing Equation~(\ref{eqa:3}).
   \STATE $C_i$ sends its encoder update $\delta_i^{t+1}$ to $S$, where $\delta_i^{t+1} = L_i^{t+1}-G^t$.
   \ELSE
   \STATE $C_i$ trains $G^t$ on $D_i$ to produce its local encoder $\tilde{L}_i^{t+1}$ with backdoor injection, by optimizing Equation~(\ref{eqa:6}).
   \STATE $C_i$ sends its encoder update $\tilde{\delta}_i^{t+1}$ to $S$, where $\tilde{\delta}_i^{t+1} = \tilde{L}_i^{t+1}-G^t$.
   \ENDIF
   \ENDFOR
   \STATE $S$ averages the received updates to generate a new global encoder 
   $G^{t+1} = G^t + \frac{\eta}{K} \sum_{i=1}^K (  \delta_i^{t+1} + \tilde{\delta}_i^{t+1}  )$, 
   where $\eta$ is a  learning rate.
   \STATE Repeat Step $5 \sim 14$ until $G^{t+1}$ get converged. 
  \end{algorithmic}
\end{algorithm}

\subsection{Backdoor Attack Process}
The backdoor attack process in FCL is presented in Algorithm~\ref{alg:backdoor}. In particular, at round $t$, attackers perform backdoor injection on local models by optimizing Equation~(\ref{eqa:6}), while benign clients normally train their local models by optimizing Equation~(\ref{eqa:3}). Next, all users upload their updates to the central server. We use $\delta^{t+1}$ and $\Tilde{\delta}^{t+1}$ to represent the updates of benign clients and attackers, respectively.

\subsection{Centralized Attack v.s. Decentralized Attack}
Considering the consistency of different attackers' objectives, we propose two typical backdoor attackers against FCL, namely the \textit{centralized attack} and the \textit{decentralized attack}.  Attackers have the same target in the centralized attack, while each attacker has its own target in the decentralized attack.

Figure~\ref{diff_attack} shows these two typical backdoor attacks against FCL, depending on whether the objectives of attackers are consistent or not. In Figure~\ref{diff_attack}, we assume there are three attackers. These attackers can collaboratively work for the same goal, or each malicious client independently poisons its own target. For simplicity, we also assume that there are three target downstream tasks. For each target downstream task, only one class is chosen as the target class. We use $(T_i,y_i)$ to represent these three \textit{(target downstream task, target class)} pairs, where $i = 1,2,3$. 

\subsubsection{Centralized attack}
In the centralized attack, the targets of all attackers are identical. All of them attempt to attack three downstream tasks. In other words, the goal of each malicious client is $\{(T_1,y_1),(T_2,y_2),(T_3,y_3)\}$. Consequently, as shown in Figure~\ref{Centralized Attack}, the updates of all attackers are similar in the parameter space. This centralized attack is analogous to the Sybils attack~\cite{douceur2002sybil, fung2020limitations}, in which a single adversary can control malicious clients to attack the system. 

\subsubsection{Decentralized attack}
In the decentralized attack, the targets of all attackers are distinct. Each attacker has its own target. Specifically, \textit{Attacker $1$} aims to
attack $\{T_1,y_1\}$, \textit{Attacker $2$} aims to attack $\{T_2,y_2\}$, and \textit{Attacker $3$} aims to attack $\{T_3,y_3\}$, respectively. Accordingly, as shown in Figure~\ref{Decentralized Attack}, the update of each attacker is diverse in the parameter space. 

\begin{table*}[!t]
  \centering
  \caption{Datasets description.}
  \begin{tabular}{ccccc} 
  \hline
  \textbf{Datasets} & \textbf{Shape} & \textbf{Classes} & \textbf{Number of training data} & \textbf{Number of test data} \\ 
  \hline
  SVHN & [3,32,32] & 10 & 73,257 & 26,032 \\ 
  CIFAR10 & [3,32,32] & 10 & 50,000 & 10,000 \\ 
  STL10 & [3,96,96] & 10 & 100,000+5,000 & 8,000 \\ 
  Animals & [3,32,32] & 10 & 19,635 & 6,544 \\ 
  GTSRB & [3,32,32] & 43 & 39,029 & 12,630 \\    
  ImageNet32 & [3,32,32] & 1,000 & 1,281,167 & 50,000 \\
  \hline
  \end{tabular}
  \label{datasets}
\end{table*}

\section{Evaluation}
\label{sec:evaluation}
\subsection{Experimental Settings}
\subsubsection{Datasets}
We conduct experiments on the SVHN~\cite{netzer2011reading}, CIFAR10~\cite{CIFAR}, STL10~\cite{coates2011analysis}, Animals~\cite{animals}, GTSRB~\cite{stallkamp2012man}, and ImageNet32~\cite{chrabaszcz2017downsampled} datasets. 

\begin{enumerate}
\item [a)] \textit{Street View House Numbers (SVHN)}~\cite{netzer2011reading} is a real-word image dataset, obtained from house numbers in Google Street View images. It contains 10 classes ranging from the digit `0' to the digit `9'. Generally, SVHN has 73,257 digits for model training, and 26,032 digits for model testing.

\item [b)] \textit{CIFAR10}~\cite{CIFAR} consists of 60,000 color images, formulated in 10 classes such as airplanes, automobiles, birds, cats, deers, dogs, frogs, horses, ships, and trucks. For each class, there are 5,000 training images and 1,000 testing images.

\item [c)] \textit{STL10}~\cite{coates2011analysis} is an image dataset derived from the \textit{ImageNet}~\cite{chrabaszcz2017downsampled}. It includes 13,000 labeled images in 10 classes such as trucks, birds, and cats, as well as 100,000 unlabeled images. Thus, it is usually used in unsupervised learning. 

\item [d)] \textit{Animals}~\cite{animals} is an animal dataset from Google images. It has 10 classes such as dogs, cats, horses, spiders, butterflies, chickens, sheep, cows, squirrels, and elephants. The data size of training and testing are 19,635 and 6,544, respectively.

\item [e)] \textit{German Traffic Sign Recognition Benchmark (GTSRB)}~\cite{stallkamp2012man} is a traffic sign dataset with different traffic backgrounds and light conditions. It covers 43 traffic signs, consisting of 39,029 training images and 12,630 testing images.

\item [f)] \textit{ImageNet32}~\cite{chrabaszcz2017downsampled} is a large dataset composed of small images known as the down-sampled version of ImageNet. It consists of 1,281,167 training images and 50,000 test images, encompassing 1,000 labels.
\end{enumerate}

The detailed dataset information in our experiment is listed in Table~\ref{datasets}. 
For consistency, we reshape the STL10 dataset to $3\times 32\times 32$. To test a poisoned encoder, we use SVHN, CIFAR10 and GTSRB as target downstream datasets, and the target class for SVHN, CIFAR10 and GTSRB is `one', `truck' and `yield sign', respectively. Unless otherwise mentioned, we use STL10 as the local training dataset of each client.

\subsubsection{Implementation details} 
We implement our attacks in python3.7 using Pytorch~\cite{paszke2017automatic}, and our experiment is evaluated on $4\times$ NVIDIA V100 GPU. 
For each experiment, we run it ten times and report the average value in this paper.
For all experiments, we elaborately pick one image as the reference data for each (\textit{target downstream dataset, target class}). Experiments results show that one standard reference data is enough to perform the attack. We set the total number of malicious clients as $3$, where each malicious client corresponds to a (\textit{target downstream dataset, target class}) pair. To quickly evaluate our proposed attack, we execute a complete attack. That is to say, $3$ malicious clients are consistently selected for the attack round, while $7$ benign clients are randomly selected. Malicious clients and benign clients form a total of $10$ selected participants for the attack round. 
The hyperparameters $\lambda_1$, $\lambda_2$ and $\lambda_3$ are set to $1$ by default. 
The local training epochs for malicious clients and benign clients are set to $10$ and $1$, respectively. 
Besides, we pretrain ResNet-50~\cite{he2016deep} with benign clients for $200$ rounds to initialize the global encoder.
To evaluate the encoder, following existing solutions, we use a weighted-KNN evaluation~\cite{wu2018unsupervised} to monitor the training process, and a linear evaluation~\cite{zhang2016colorful,van2018representation,bachman2019learning} to test the final performance.

\subsubsection{Encoder Evaluation} 
There are two common methods to evaluate an encoder, namely linear evaluation~\cite{zhang2016colorful, van2018representation, bachman2019learning} and weighted-KNN evaluation~\cite{wu2018unsupervised}. 
Linear evaluation trains a linear model on the feature representations extracted from the frozen encoder. To classify a sample $x$, the weighted-KNN evaluation extracts the feature representations of all test samples from the frozen encoder. Then, the weighted-KNN evaluation compares the feature representations of $x$ against those of all test samples, using cosine similarity. The top $k$ nearest neighbors are used to make the prediction via weighted voting. Since the computation cost is prohibitive for training a linear classifier at every communication round, we adopt weighted-KNN evaluation to monitor the training process, and linear evaluation to test the final performance. 

\subsubsection{Multi-shot scenario v.s. one-shot scenario} 
We evaluate our proposed attacks in two common scenarios, namely the \textit{multi-shot} scenario and the \textit{one-shot} scenario~\cite{bagdasaryan2020backdoor,xie2019dba}. In the \textit{multi-shot} scenario, malicious clients perform backdoor attacks for local models and then upload the poisoned updates to further poison the global model. Malicious clients need to be consistently selected as participating clients for multiple rounds, otherwise, the backdoor will be forgotten quickly by the global model. In the \textit{one-shot} scenario, malicious clients perform backdoor attacks for local models, scale the poisoned update, and then upload the scaled update to further poison the global model. The scaling operation suppresses the benign update, which makes the poisoned update survive during the global model aggregation. Thus, it only needs one round attack in the one-shot scenario.

\begin{figure*}[h]
    \centering
    \subfigure[SVHN + IID data. ] 
    {\includegraphics[width=0.31\textwidth]{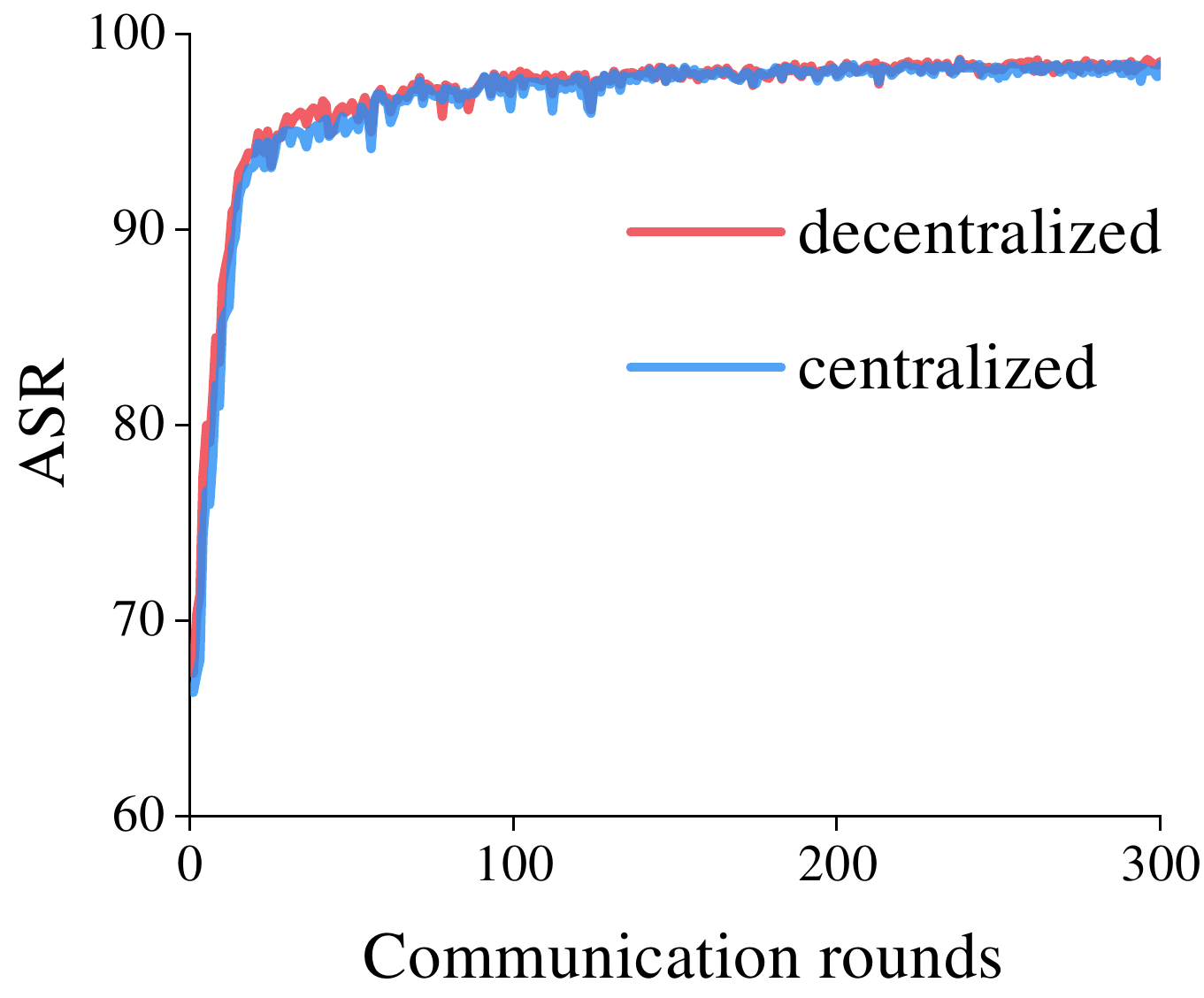}} 
    \subfigure[CIFAR10 + IID data.]
    {\includegraphics[width=0.31\textwidth]{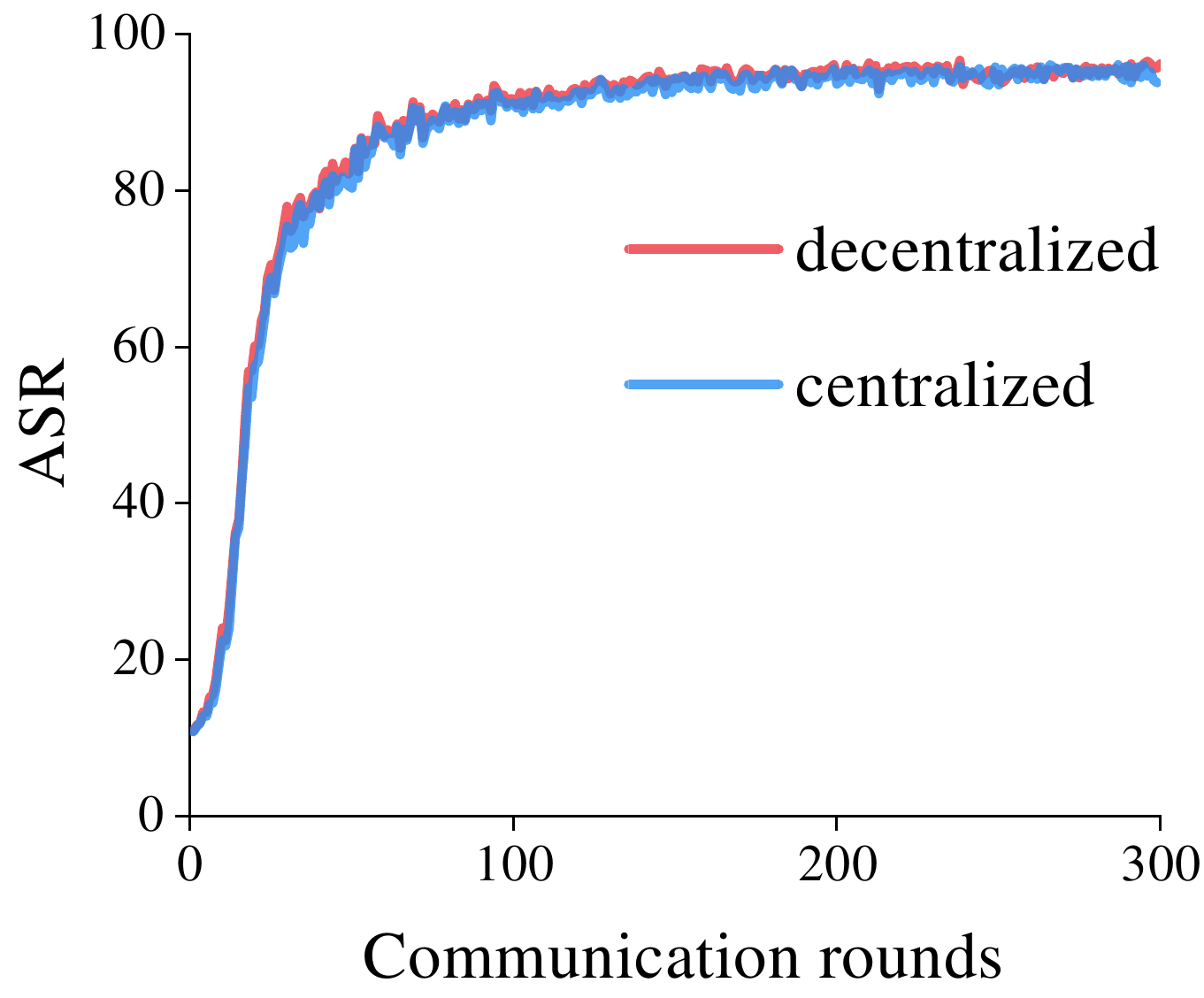}} 
    \subfigure[GTSRB + IID data.]
    {\includegraphics[width=0.31\textwidth]{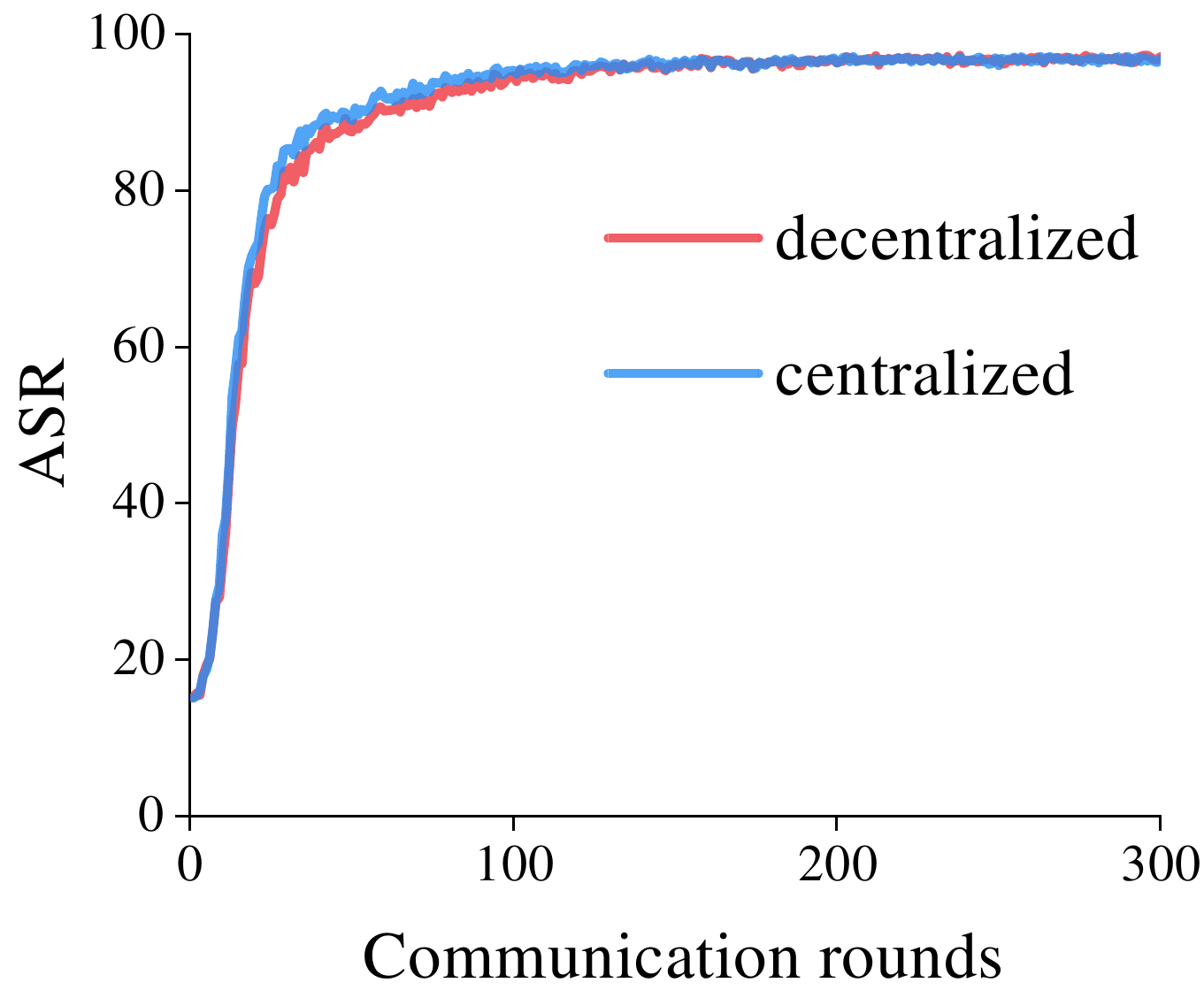}}
    \\
    \subfigure[SVHN + non-IID data.]
    {\includegraphics[width=0.31\textwidth]{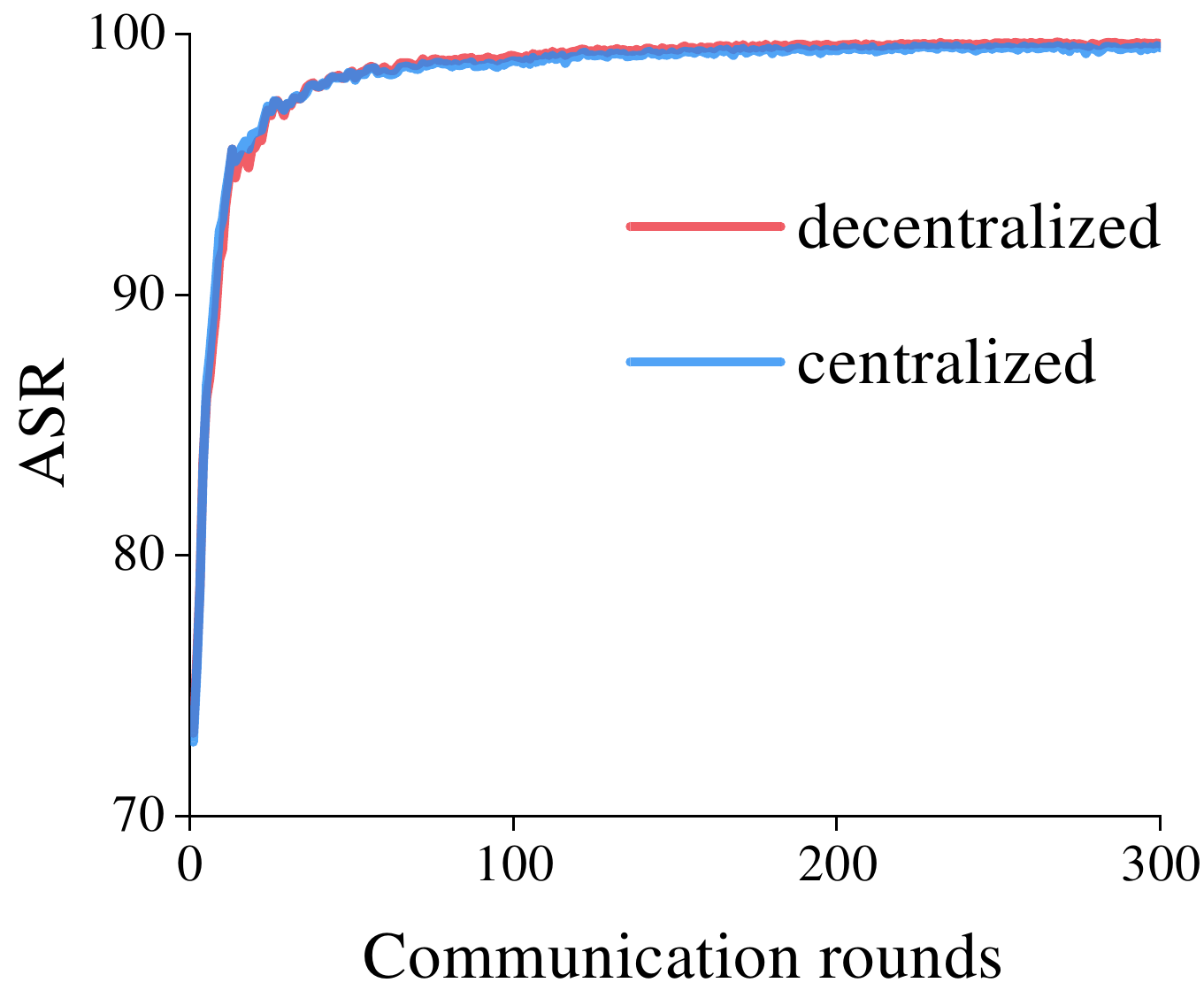}} 
    \subfigure[CIFAR10 + non-IID data.]
    {\includegraphics[width=0.31\textwidth]{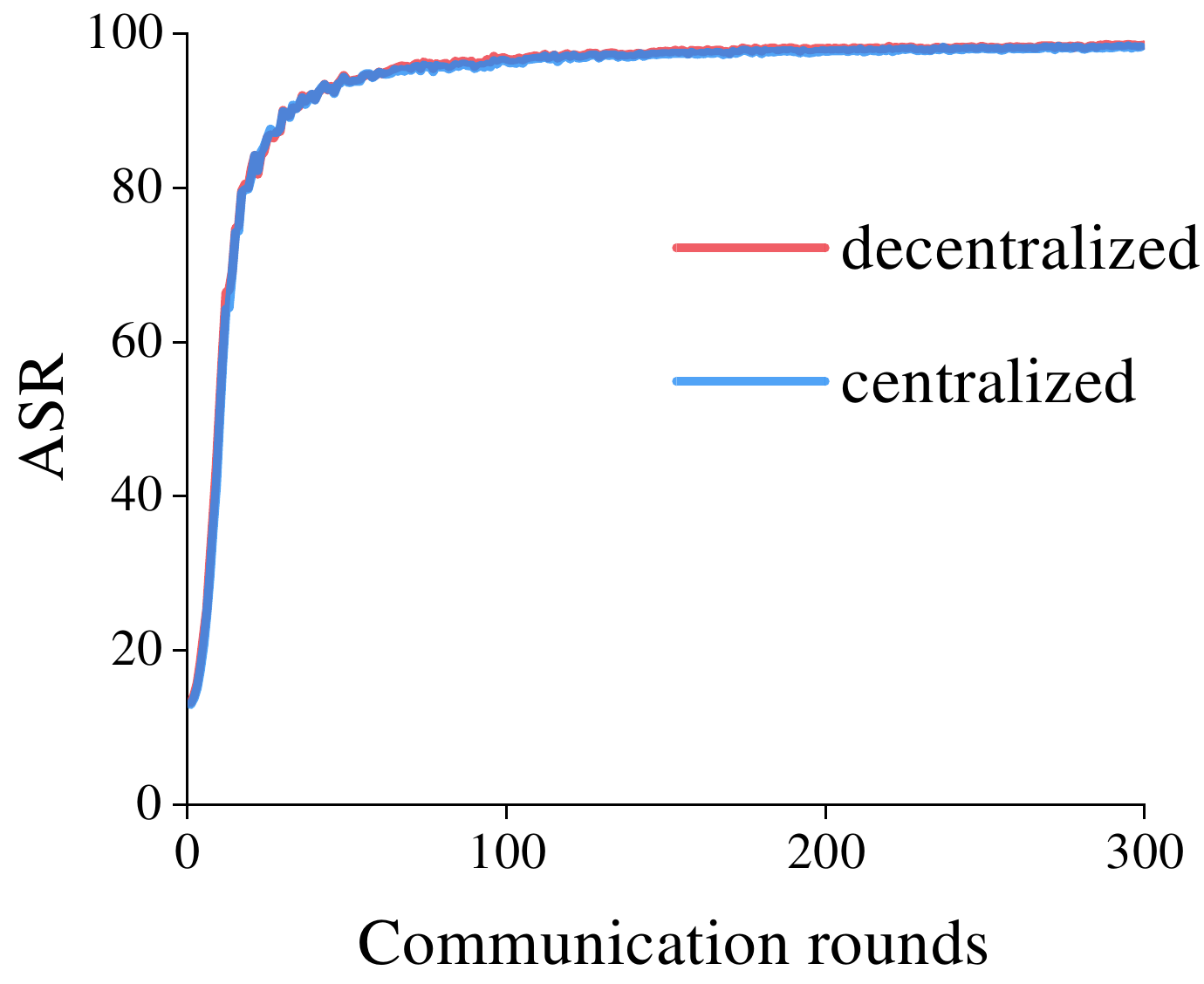}} 
    \subfigure[GTSRB + non-IID data.]
    {\includegraphics[width=0.31\textwidth]{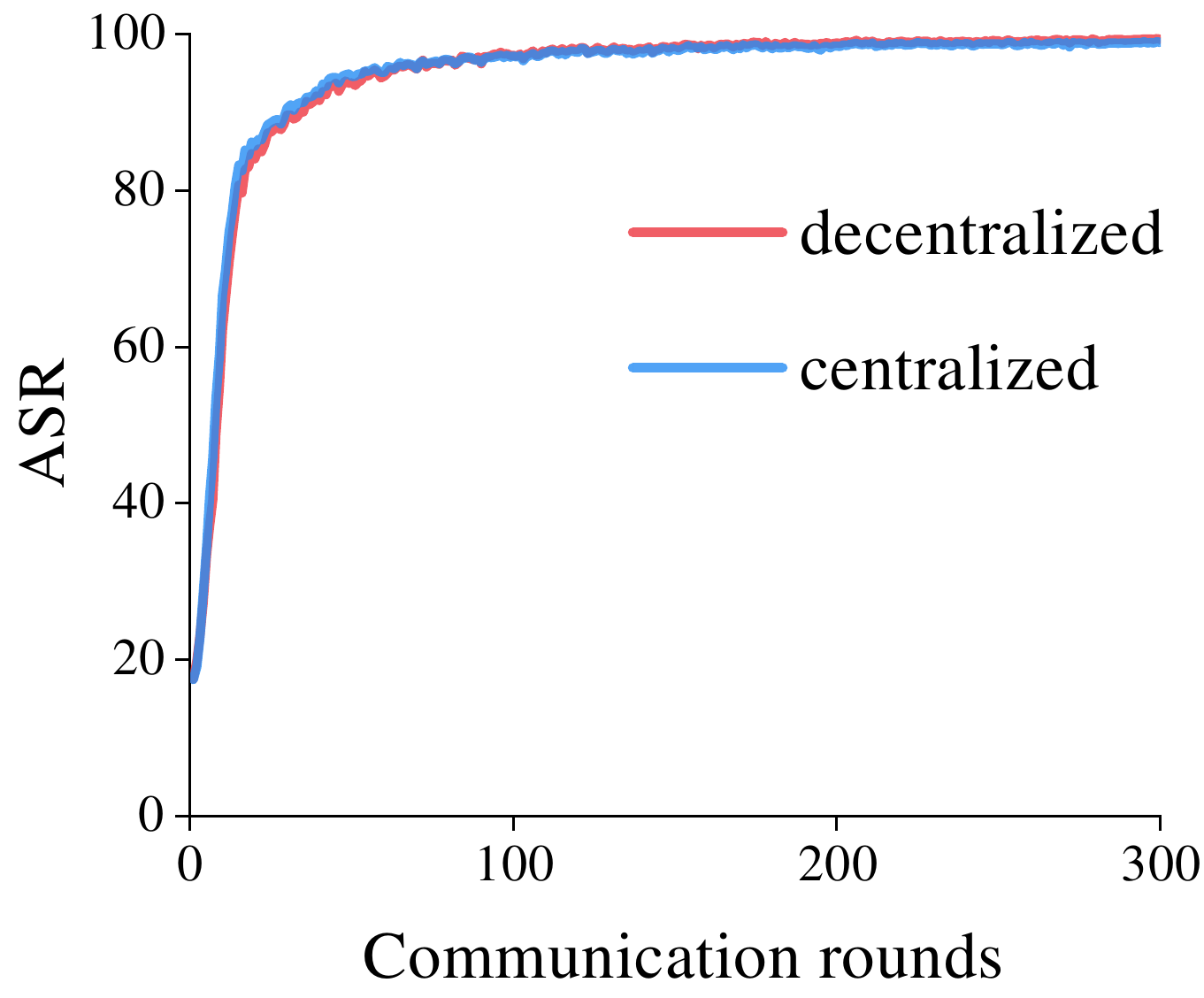}}
    \\
    \caption{ASRs in the multi-shot scenario on both IID and non-IID data.}
    \label{iid_noniid}
\end{figure*}

\begin{table*}[h]
  \centering
  \caption{Evaluation results in the multi-shot scenario on both IID and non-IID data.}
    \begin{tabular}{cccccccc} 
    \hline
    \multirow{2}{*}{\begin{tabular}[c]{@{}c@{}}\textbf{Data distribution }\end{tabular}} & \multirow{2}{*}{\begin{tabular}[c]{@{}c@{}}\textbf{Downstream datasets}\end{tabular}} & \multicolumn{2}{c}{\begin{tabular}[c]{@{}c@{}}\textbf{Centralized attack}\end{tabular}} & \multicolumn{2}{c}{\begin{tabular}[c]{@{}c@{}}\textbf{Decentralized attack}\end{tabular}} & \multicolumn{2}{c}{\begin{tabular}[c]{@{}c@{}}\textbf{Baseline(no attack)}\end{tabular}} \\ 
    \cline{3-8}
     &  & \textit{main acc(\%)} & \textit{ASR(\%)} & \textit{main acc(\%)} & \textit{ASR(\%)} & \textit{main acc(\%)} & \textit{ASR(\%)} \\ 
    \hline
    \multirow{3}{*}{IID} & SVHN & 57.34 & 78.49 & 56.88 & {\cellcolor[rgb]{0.902,0.902,0.902}}\textbf{90.44} & 61.28 & 30.87 \\
     & CIFAR10 & 77.54 & 93.28 & 76.94 & {\cellcolor[rgb]{0.902,0.902,0.902}}\textbf{96.05} & 77.07 & 11.58 \\
     & GTSRB & 88.25 & 95.54 & 88.28 & {\cellcolor[rgb]{0.902,0.902,0.902}}\textbf{95.77} & 88.72 & 18.01 \\ 
    \hline
    \multirow{3}{*}{non-IID} & SVHN & 57.14 & 74.82 & 56.69 & {\cellcolor[rgb]{0.902,0.902,0.902}}\textbf{82.76} & 61.54 & 35.57 \\
     & CIFAR10 & 76.70 & 88.95 & 77.36 & {\cellcolor[rgb]{0.902,0.902,0.902}}\textbf{91.23} & 77.68 & 11.42 \\
     & GTSRB & 87.04 & 95.13 & 88.45 & {\cellcolor[rgb]{0.902,0.902,0.902}}\textbf{95.17} & 88.63 & 19.48 \\
    \hline
    \end{tabular}
  \label{linear_evaluation}
  \end{table*}

\subsubsection{Evaluation Metrics} 
We have two evaluation metrics for our attacks, namely \textit{Main Accuracy (main acc)} and \textit{Attack Success Rate (ASR)}. 
For a given downstream task, the \textit{main acc} is the test accuracy of the poisoned downstream classifier on clean test images. For a backdoor attack in FCL, it needs to maintain high performance on clean test images.
For a given downstream task and a target class, 
the \textit{ASR} is determined by the poisoned downstream model. It means the percentage of trigger-embedded test images predicted as the target class. A higher value of \textit{ASR} indicates a more successful attack.

\subsection{Experiment Results}
\subsubsection{Effectiveness in the multi-shot scenario}
In this experiment, we evaluate our centralized and decentralized backdoor attacks against FCL in a multi-shot scenario on both IID and non-IID data. Figure~\ref{iid_noniid} shows the \textit{ASR}s during the training process. We can infer that the \textit{ASR}s of both attacks increase with the communication round, and also perform well among three target downstream datasets on both IID and non-IID data. 

Table~\ref{linear_evaluation} shows the \textit{ASR}s of the final global encoder. The baseline is evaluated on the encoder which is trained by benign clients,  excluding malicious clients. As we can see, our attacks achieve higher \textit{ASR}s than the baseline. Also, the \textit{ASR}s of the decentralized attack is higher than that of the centralized attack in all cases.
Finally, both attacks perform better on IID data than non-IID data. 

\begin{figure*}[htbp]
\centering
\subfigure[SVHN.]{
\includegraphics[width=0.31\linewidth]{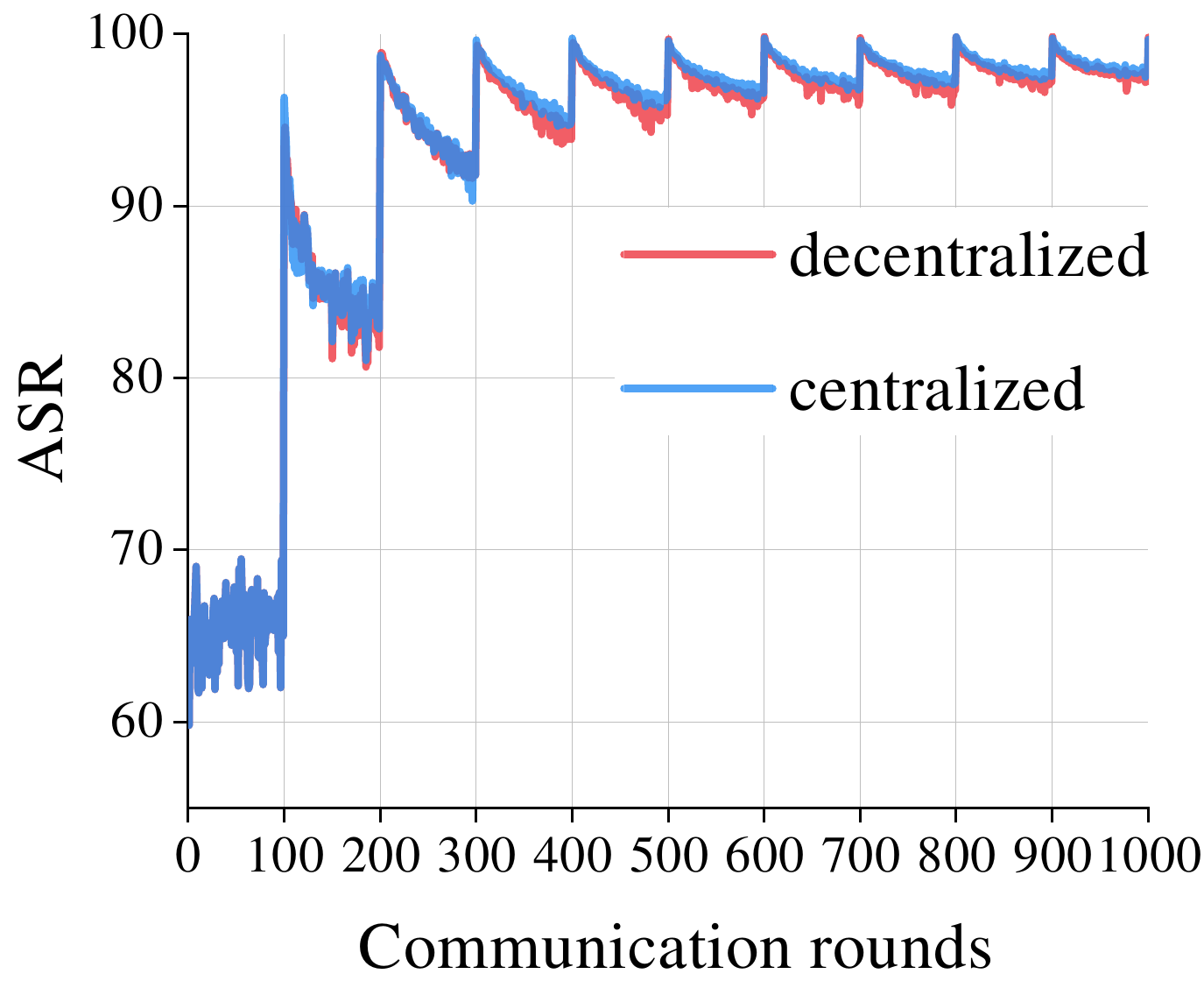} 
}
\subfigure[CIFAR10.]{
\includegraphics[width=0.31\linewidth]{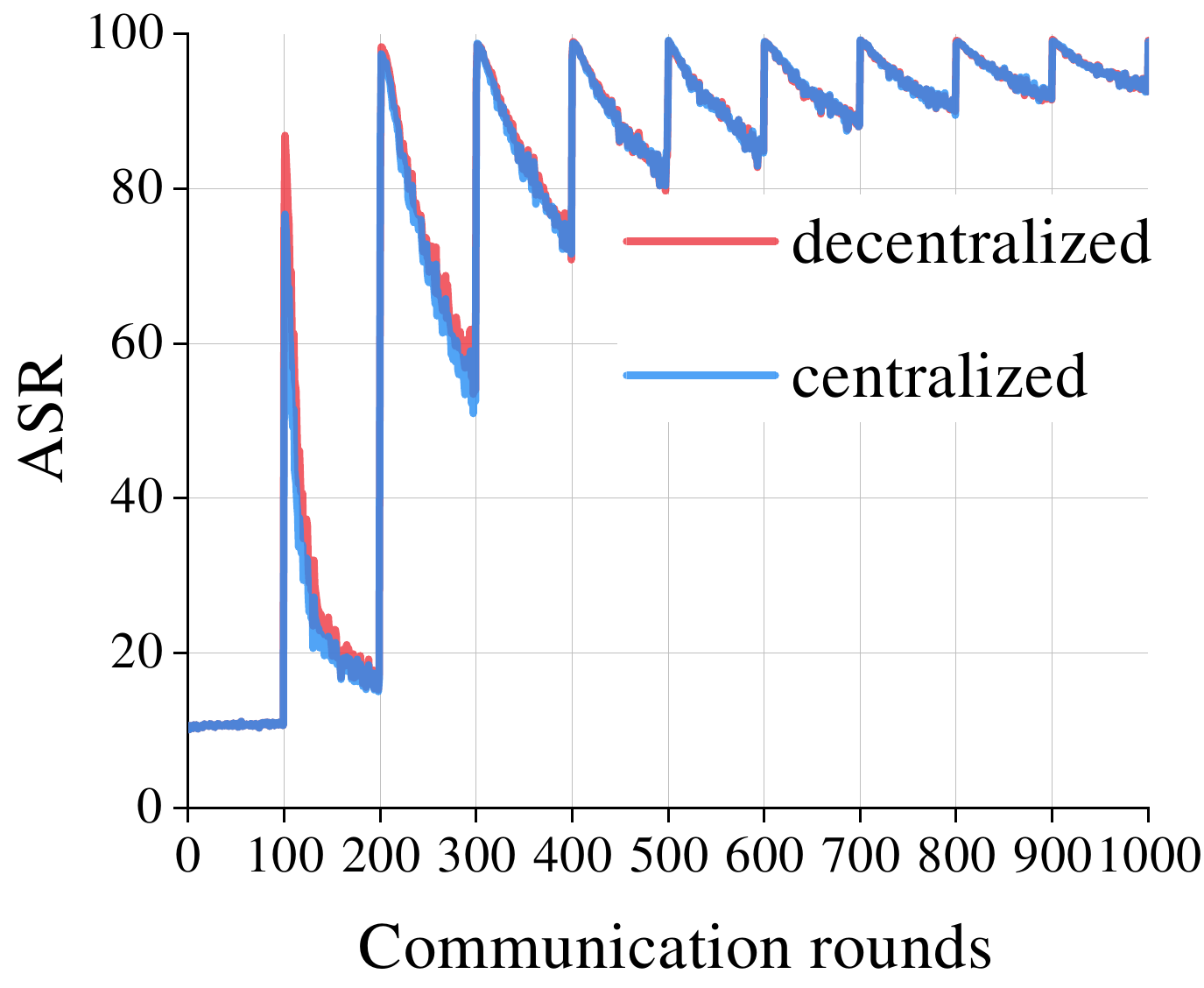} 
}
\subfigure[GTSRB.]{
\includegraphics[width=0.31\linewidth]{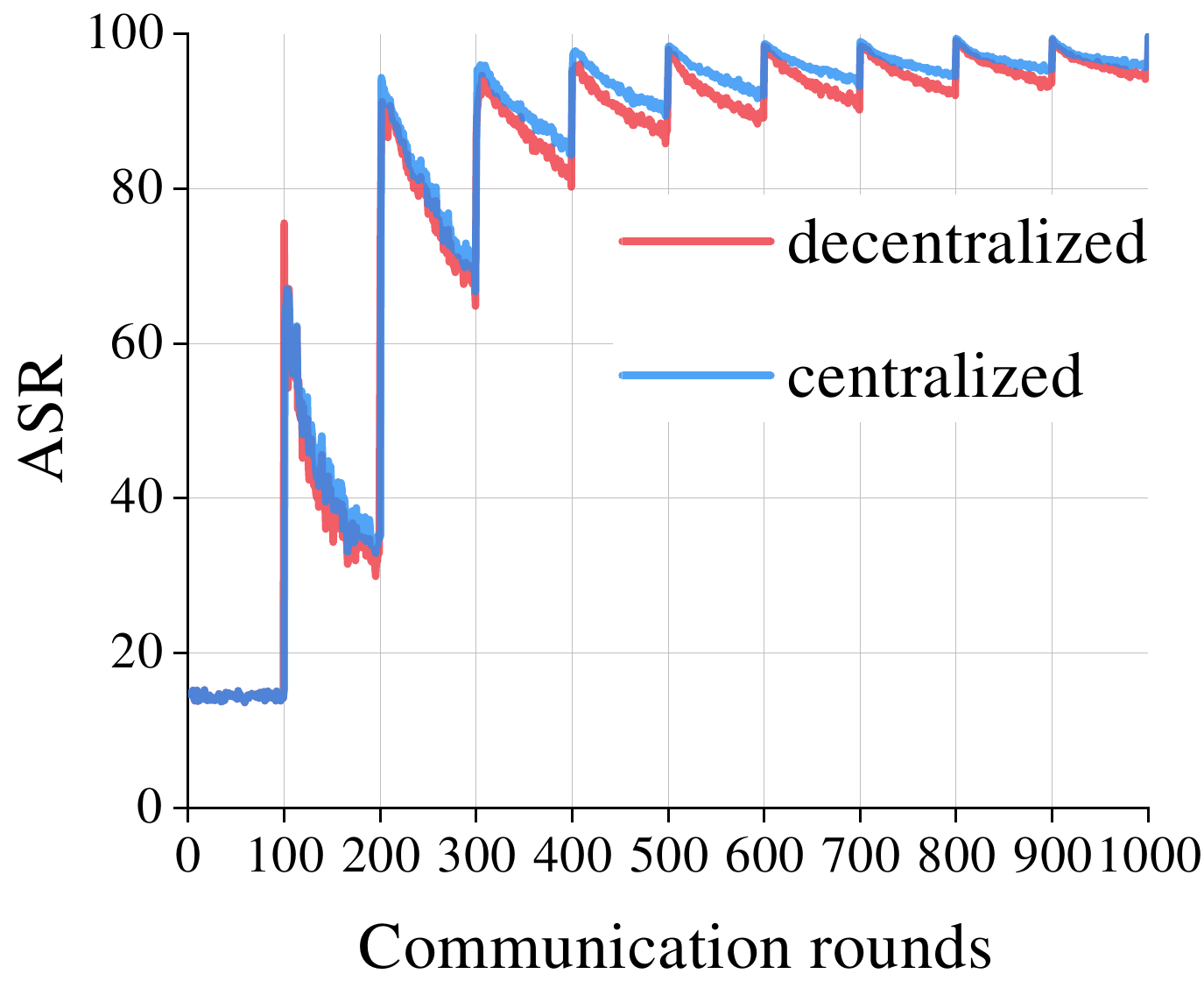}
}
\caption{ASRs in the one-shot scenario on IID data.}
\label{one_shot}
\end{figure*}

\begin{figure}[!t]
  \centering
  \includegraphics[width=0.36\textwidth]{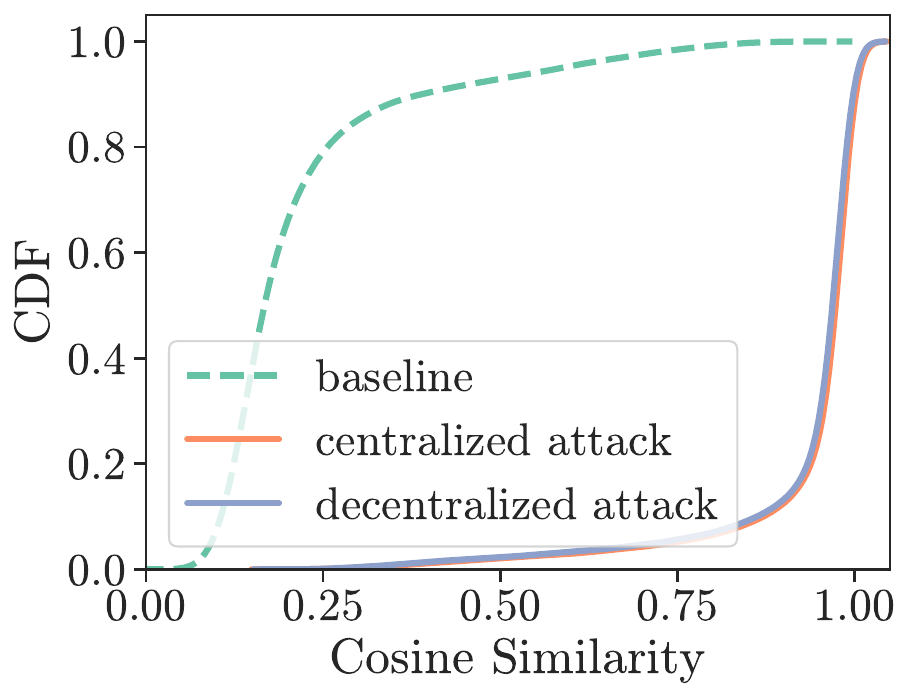}
  \caption{The cumulative distribution functions (CDFs) of cosine similarities between the features of reference data and that of trigger-embedded data (baseline is clean data) generated by the backdoored encoder.}
  \label{Cos_sim}
\end{figure}

\begin{table*}[h]
  \centering
  \caption{Different training data of malicious clients.}
    \begin{tabular}{cccccc} 
      \hline
      \multirow{2}{*}{\begin{tabular}[c]{@{}c@{}}\textbf{Training data}\\\textbf{of malicious clients}\end{tabular}} & \multirow{2}{*}{\textbf{Downstream datasets }} & \multicolumn{2}{c}{\textbf{~Centralized attack~}} & \multicolumn{2}{c}{\textbf{Decentralized attack}} \\ 
      \cline{3-6}
       &  & \textit{main acc(\%)} & \textit{ASR} & \textit{main acc(\%)} & \textit{ASR} \\ 
      \hline
      \multirow{3}{*}{Animals} & SVHN & 57.27 & {\cellcolor[rgb]{0.894,0.894,0.894}}86.26 & 58.83 & {\cellcolor[rgb]{0.894,0.894,0.894}}90.08~ \\
       & CIFAR10 & 77.25 & {\cellcolor[rgb]{0.894,0.894,0.894}}46.73 & 77.64 & {\cellcolor[rgb]{0.894,0.894,0.894}}63.23 \\
       & GTSRB & 86.34 & {\cellcolor[rgb]{0.894,0.894,0.894}}87.57 & 88.55 & {\cellcolor[rgb]{0.894,0.894,0.894}}88.71 \\ 
      \hline
      \multirow{3}{*}{STL10} & SVHN & 58.29 & {\cellcolor[rgb]{0.784,0.784,0.784}}90.51 & 58.11 & {\cellcolor[rgb]{0.784,0.784,0.784}}92.44 \\
       & CIFAR10 & 76.48 & {\cellcolor[rgb]{0.784,0.784,0.784}}52.22 & 77.72 & {\cellcolor[rgb]{0.784,0.784,0.784}}67.96 \\
       & GTSRB & 87.91 & {\cellcolor[rgb]{0.784,0.784,0.784}}94.09 & 87.13 & {\cellcolor[rgb]{0.784,0.784,0.784}}95.49 \\ 
      \hline
      \multirow{3}{*}{ImageNet32} & SVHN & 58.30 & {\cellcolor[rgb]{0.6,0.6,0.6}}95.35 & 57.12 & {\cellcolor[rgb]{0.6,0.6,0.6}}96.21 \\
       & CIFAR10 & 77.76 & {\cellcolor[rgb]{0.6,0.6,0.6}}92.17 & 78.37 & {\cellcolor[rgb]{0.6,0.6,0.6}}95.85 \\
       & GTSRB & 88.15 & {\cellcolor[rgb]{0.6,0.6,0.6}}95.04 & 88.44 & {\cellcolor[rgb]{0.6,0.6,0.6}}96.60 \\
      \hline
      \end{tabular}                                        
  \label{influence_dataset}
\end{table*}

To further explore the backdoor injection status, we compute the feature representations of the reference data and the trigger-embedded test data. Then, we plot the cumulative distribution functions (CDFs) of the cosine similarity values of these feature representations.
As shown in Figure~\ref{Cos_sim}, most of the cosine similarity values are in close proximity to $1.0$ for both the centralized and the decentralized attacks, significantly different from the baseline (i.e., clean data). It indicates that the global encoder in our attacks successfully makes the trigger-embedded test data produce very similar feature representations to the reference data.

By summarizing the above results, we can see that the centralized and the decentralized attacks are effective in multi-shot scenarios. Both of them could backdoor the global encoder. Compared with the centralized attack, the decentralized attack is more powerful in general. Moreover, the attack performance could be affected by data distribution.

\subsubsection{Effectiveness in the one-shot scenario} 
In this experiment, we evaluate our backdoor attacks against FCL in a one-shot scenario on IID data.
We set the scale factor $\gamma=100$, and the attack frequency as $1/100$ (i.e., attack once for every 100 communication rounds). As shown in Figure~\ref{one_shot}, the \textit{ASR}s of both attacks rise to high values suddenly when the attack round occurs, and then it gradually decreases until the next attack round comes. This phenomenon indicates that our two attacks perform well in this one-shot scenario.

\begin{figure*}[!t]
    \centering

    \subfigure[SVHN.]{
    \includegraphics[width=0.28\textwidth]{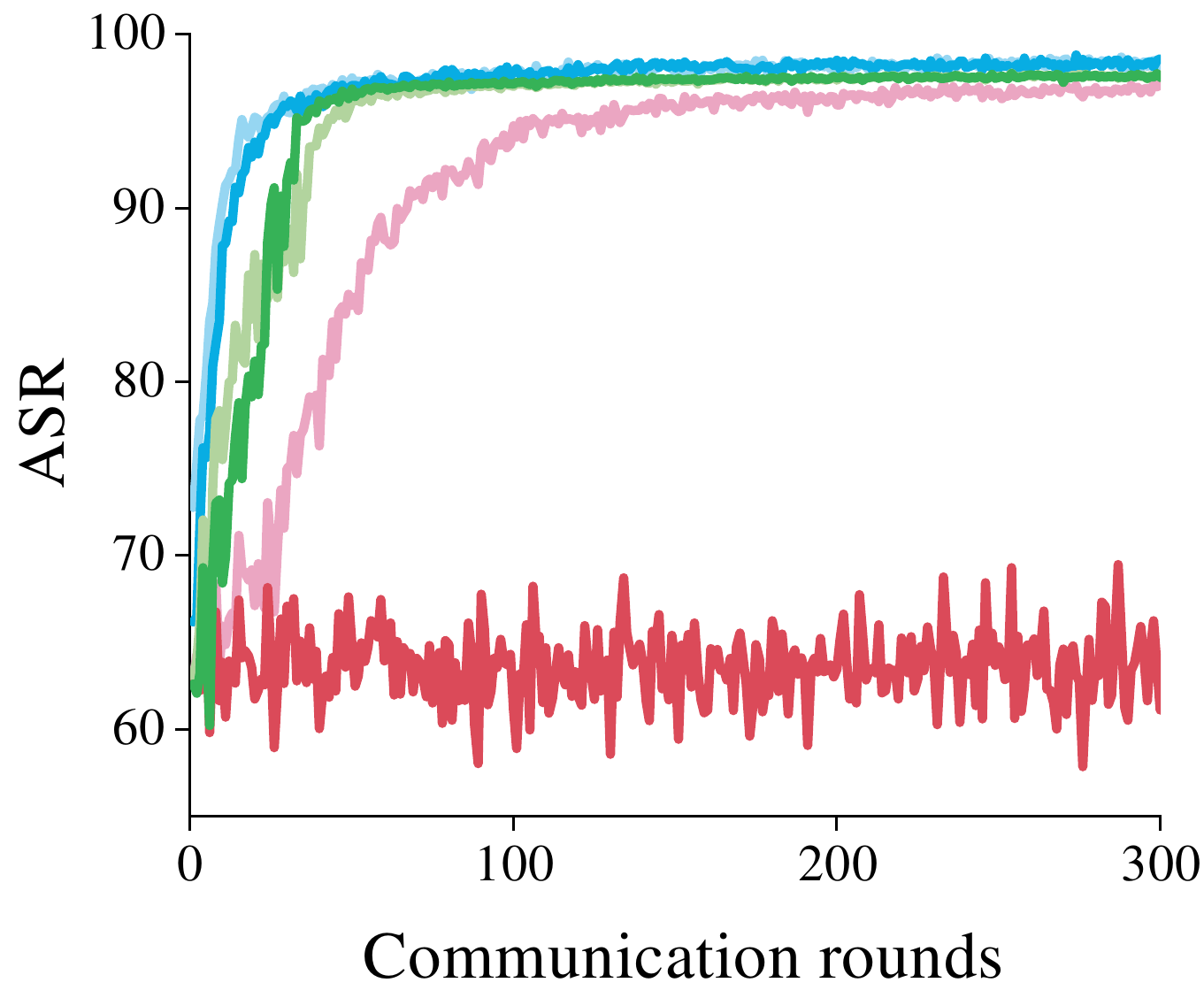}} 
    \subfigure[CIFAR10.]{
    \includegraphics[width=0.28\textwidth]{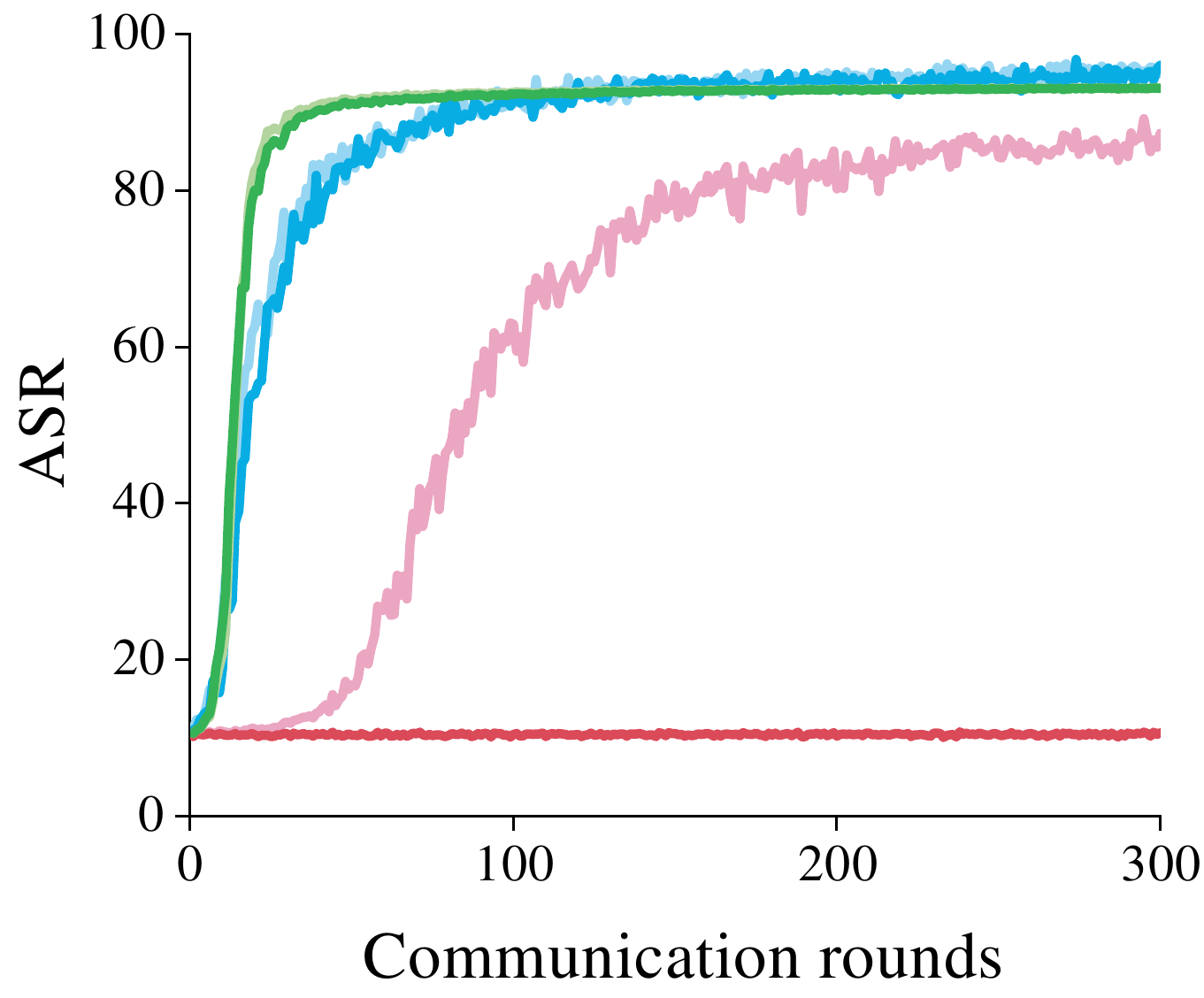}} 
    \subfigure[GTSRB.]{
    \includegraphics[width=0.28\textwidth]{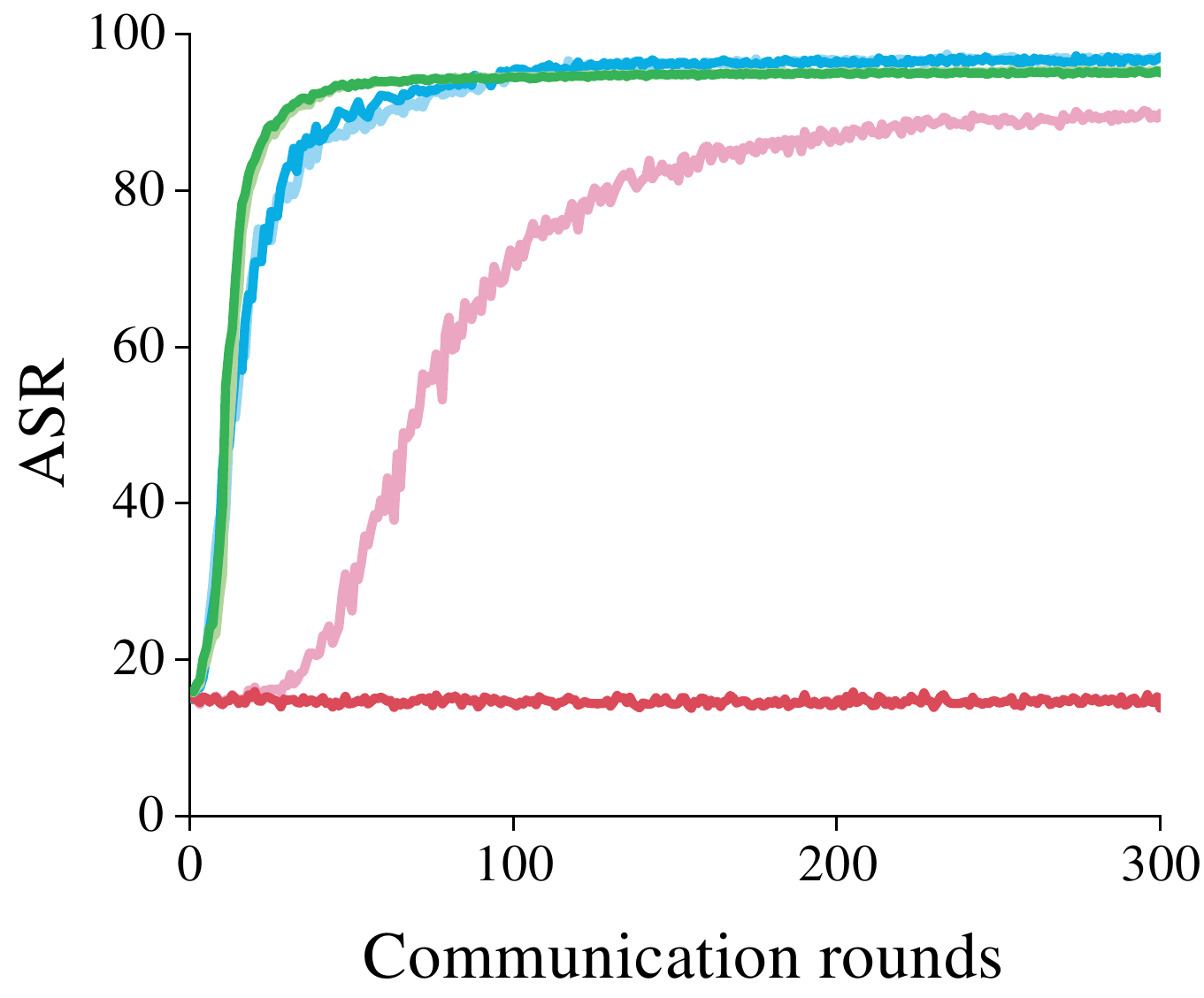}} 
    \includegraphics[width=0.12\textwidth]{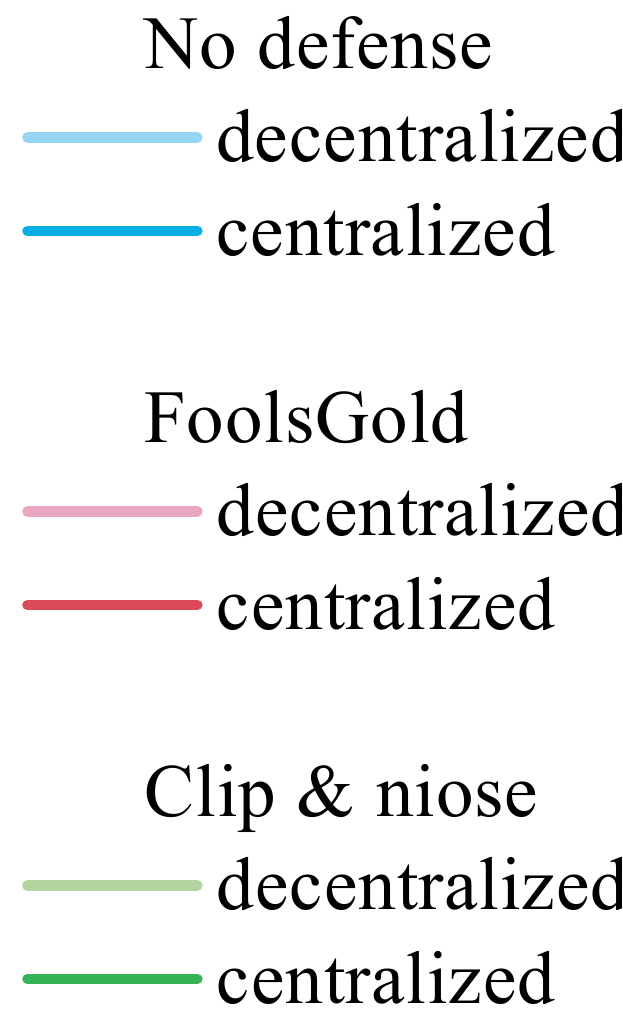}
    \\
    \caption{\textit{ASR}s under FoolsGold and Clip \& Noise defense methods in the multi-shot scenario.}
    \label{defense_ms}
\end{figure*}

\begin{figure*}[!t]
    \centering
    \subfigure[SVHN.]{
    \includegraphics[width=0.28\textwidth]{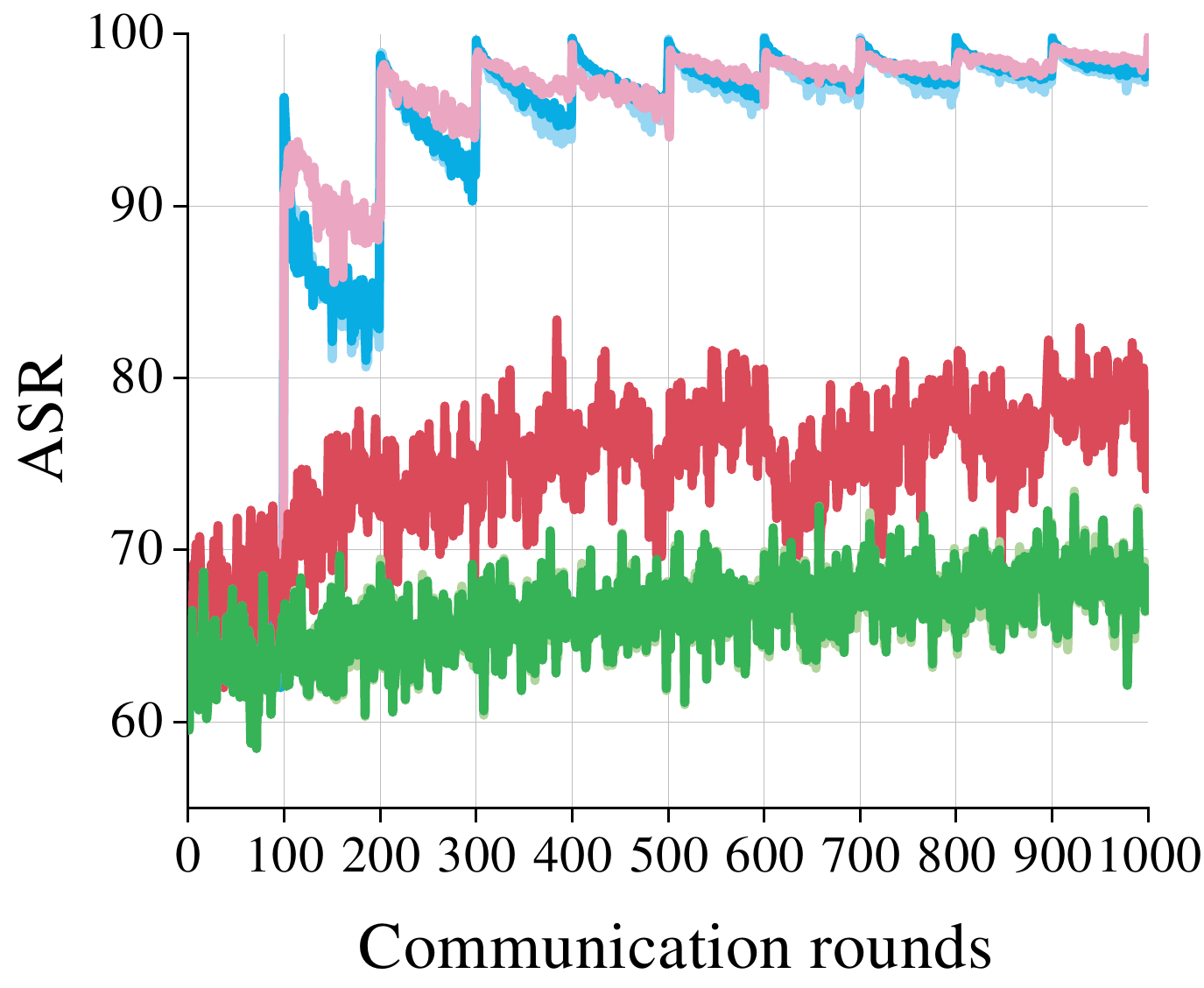}} 
    \subfigure[CIFAR10.]{
    \includegraphics[width=0.28\textwidth]{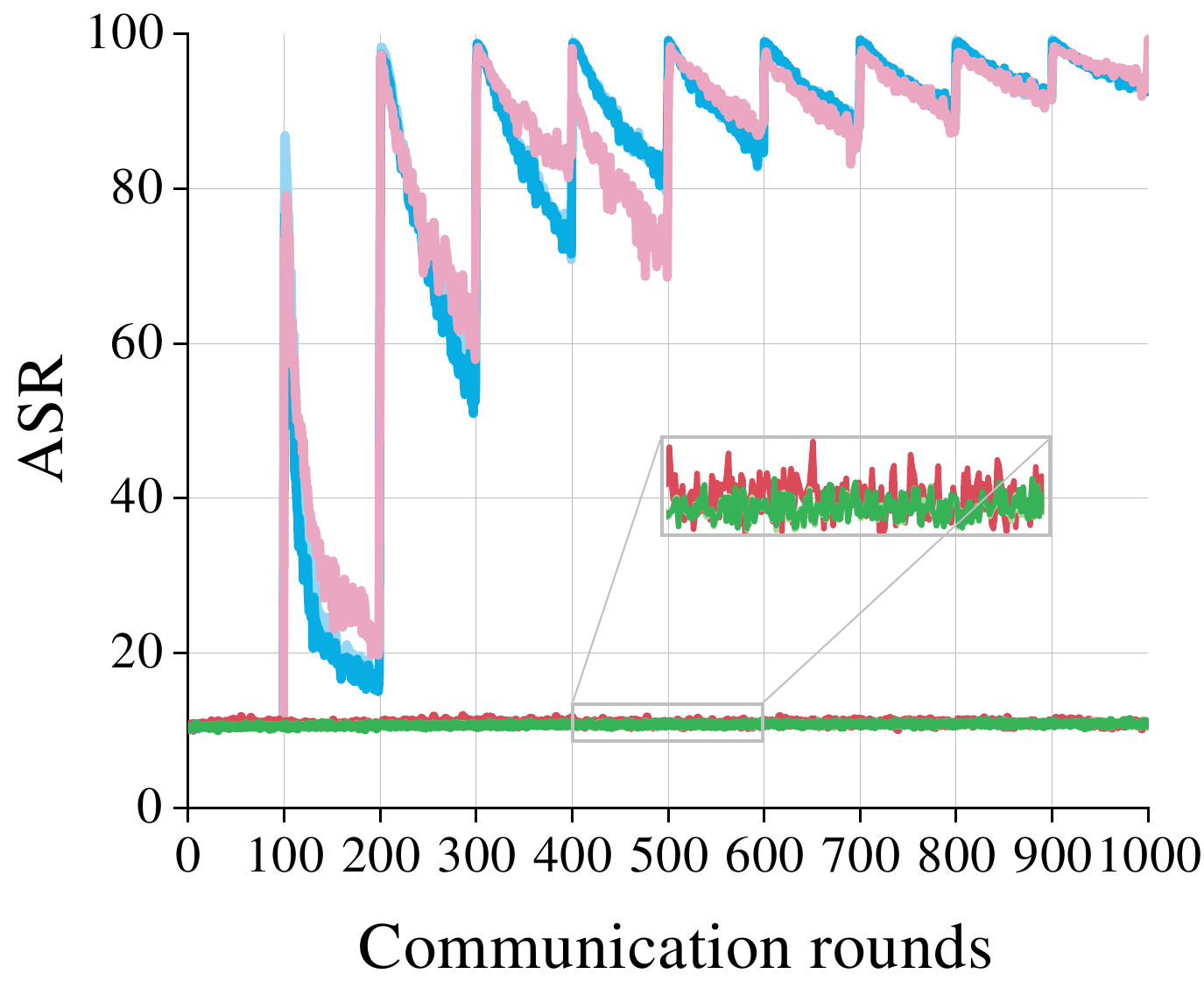}} 
    \subfigure[GTSRB.]{
    \includegraphics[width=0.28\textwidth]{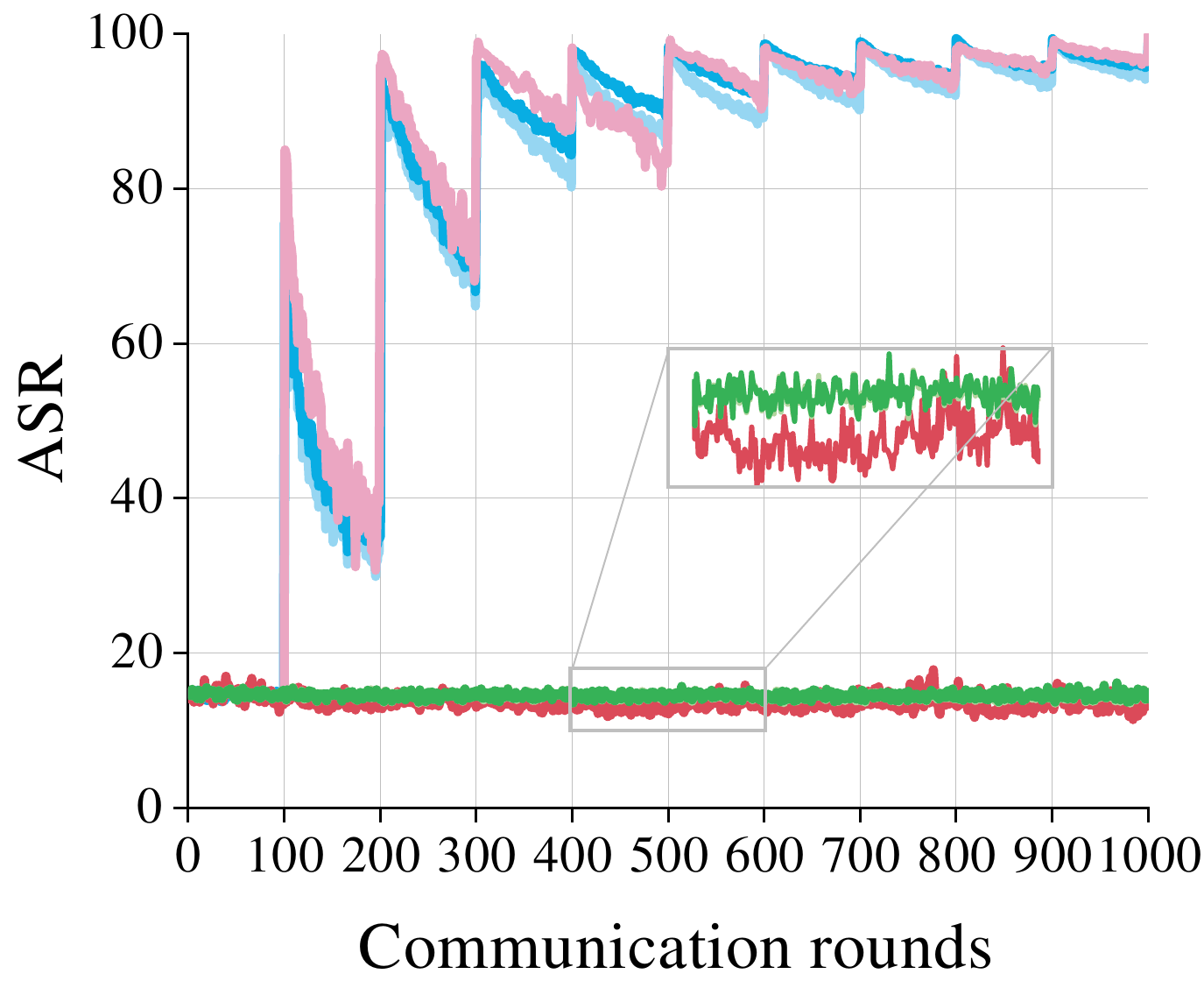}} 
    \includegraphics[width=0.12\textwidth]{figure/defense_legend.pdf}
    \\
    \caption{\textit{ASR}s under FoolsGold and Clip \& Noise defense methods in the one-shot scenario.}
    \label{defense_os}
\end{figure*}

\subsubsection{Impact of attackers' local training data}
In realistic applications, malicious clients might not have prior knowledge of benign clients' data. Thus, they usually randomly use other data (e.g., online open-source data) as their local training data. To simulate the impact of attackers' local training data, we use CIFAR10 as the training data for benign clients, while Animals, STL10, and ImageNet32 as the training data of malicious clients. 
The evaluation results are shown in Table~\ref{influence_dataset}. 
For both the centralized and the decentralized attacks, ImageNet32 produces higher \textit{ASR}s than STL10 and Animals. One possible reason is that ImageNet32 is a huge hierarchical image dataset, which is more likely to have similar data with benign clients.

\section{Defense Results}
\label{sec:defense}
We consider two typical defense methods, namely \textit{FoolsGold} and \textit{Clip \& Noise}, for our proposed backdoor attacks against FCL. 
\textit{FoolsGold} attempts to penalize the aggregation weights of participating clients who constantly produce similar updates, and retain the aggregation weights of clients who contribute dissimilar updates~~\cite{fung2020limitations}. Thus, FoolsGold can be used to mitigate the sybil attack~\cite{douceur2002sybil, fung2020limitations}.
\textit{Clip \& Noise} contains weight clipping and noise addition in the model aggregation. Specifically, in the central server, any client update, whose $L_2$ norm exceeds a given threshold, will be clipped. Then, all client updates will be biased with a Gaussian noise for further weighted aggregation~\cite{sun2019can}.

Figure~\ref{defense_ms} shows the defense results in the multi-shot scenario. We can infer that both the centralized and the decentralized attacks achieve high \textit{ASR}s under Clip \& Noise, which indicates that they are immune to Clip \& Noise. 
For FoolsGold, the \textit{ASR}s of the decentralized attack rise slowly in the first few rounds, and then gradually increase to a high value. 
In addition, the \textit{ASR}s of the centralized attack almost remain constant at a low value for FoolsGold. 
The potential reasons are as follows. In the centralized attack, malicious clients have similar encoder updates, leading to aggregation weight penalization. In the decentralized attack, malicious clients have distinct encoder updates, to escape from aggregation weight penalization in FoolsGold.

Figure~\ref{defense_os} demonstrates the defense results in the one-shot scenario. 
For Clip \& Noise, both the centralized and the decentralized attacks perform poorly. The reason is that the norm of attackers' scaled updates usually exceeds the given threshold in the one-shot scenario, which will be clipped by the server.
For FoolsGold, the decentralized attack also performs better than the centralized attack. The reason is the same with the multi-shot scenario. That is to say, the scaled updates between malicious clients are still very similar in the centralized attack, leading to aggregation weight penalization.

\section{Conclusion and Future Work}
\label{sec:conclusion}
In this paper, we investigate the security threats in FCL as a pioneer work. Specifically, we study two different backdoor attacks against FCL, namely the centralized attack and the decentralized attack. Experiments results on different datasets show that our proposed attacks can be successfully inject backdoor triggers into the global encoder as well as downstream models. We also show that the decentralized attack is more stealthy than the centralized attack. In practical terms, our work can be easily applied to different application areas. For example, in the intelligent transport area, it could mislead downstream models to recognize the `yield' sign as the `speed up' sign.

Generally, we attempt to broaden the research view of the backdoor attack against FCL. There are still many open issues. 
For example, we should consider the backdoor attack against FCL under different local training speeds or different upload bandwidths. 
We should also evaluate the security issues with different ways to build downstream models 
(e.g., in a fine-tuned manner). 
We believe these directions will further extend our study and we will leave them as future work.

\section{Acknowledgements}
This work is supported by National Natural Science Foundation of China (No. 61802383), Research Project of Pazhou Lab for Excellent Young Scholars (No. PZL2021KF0024), and Guangzhou Basic and Applied Basic Research Foundation (No. 202201010330, No. 202201020162).



\begin{IEEEbiographynophoto}{Yao~Huang}
is a master student at Institute of Artificial Intelligence and Blockchain, Guangzhou University, China. His main research interests are deep learning and security \& privacy.
\end{IEEEbiographynophoto}

\begin{IEEEbiographynophoto}{Kongyang~Chen}
is an associate professor with the Institute of Artificial Intelligence and Blockchain, Guangzhou University, China. He received his PhD degree in computer science from the University of Chinese Academy of Sciences, China. His main research interests are artificial intelligence, edge computing, distributed systems, etc. He has published more than 20 papers on top-tier conferences and journals, such as INFOCOM, ACM MobiSys, IEEE TMC, IEEE TDSC, IEEE TPDS, and ACM TOSN.
\end{IEEEbiographynophoto}

\begin{IEEEbiographynophoto}{Jiannong~Cao}
is currently the Otto Poon Charitable Foundation Professor in Data Science and the Chair Professor of Distributed and Mobile Computing in the Department of Computing at The Hong Kong Polytechnic University (PolyU), Hong Kong. He is also the Dean of Graduate School, the director of Research Institute for Artificial Intelligence of Things (RIAIoT) in PolyU, the director of the Internet and Mobile Computing Lab (IMCL). He was the founding director and now the associate director of PolyU's University's Research Facility in Big Data Analytics (UBDA). 
His research interests include distributed systems and blockchain, wireless sensing and networking, big data and machine learning, and mobile cloud and edge computing. He has served the Chair of the Technical Committee on Distributed Computing of IEEE Computer Society 2012-2014, a member of IEEE Fellows Evaluation Committee of the Computer Society and the Reliability Society, a member of IEEE Computer Society Education Awards Selection Committee, a member of IEEE Communications Society Awards Committee, and a member of Steering Committee of IEEE Transactions on Mobile Computing. He has also served as chairs and members of organizing and technical committees of many international conferences, including IEEE INFOCOM, IEEE PERCOM, IEEE IoTDI, IEEE ICPADS, IEEE CLOUDCOM, SRDS and OPODIS, and as associate editor and member of the editorial boards of many international journals, including IEEE TC, IEEE TPDS, IEEE TBD, IEEE IoT Journal, ACM ToSN, ACM TIST, ACM TCPS.
He is a member of Academia Europaea, a fellow of the Hong Kong Academy of Engineering Science, a fellow of IEEE, a fellow of China Computer Federation (CCF) and an ACM distinguished member.
\end{IEEEbiographynophoto}

\begin{IEEEbiographynophoto}{Jiaxing~Shen}
is an assistant professor with the Department of Computing and Decision Sciences at Lingnan University. Previously, he was with the Department of Computing at the Hong Kong Polytechnic University as a Research Assistant Professor from 2020 to 2022. He obtained a Ph.D. degree in Computing from the Hong Kong Polytechnic University in 2019. In 2017, he was a visiting scholar with the Media Lab at MIT. His research interests include Context Sensing, IoT Systems, Mobile Computing, and Data Mining. He has published over 30 papers including top-tier journals and conferences such as IEEE TMC, IEEE TKDE, ACM TOIS, ACM IMWUT, IEEE INFOCOM, and WWW. He has also won two Best Paper awards including one from IEEE INFOCOM 2020.
\end{IEEEbiographynophoto}

\begin{IEEEbiographynophoto}{Shaowei~Wang}
received the PhD degree in the School of Computer Science and Technology from the University of Science and Technology of China (USTC), in 2019. He is an associate professor in Institute of Artificial Intelligence and Blockchain with Guangzhou University. His research interests are data privacy, federated learning and recommendation systems. He has published more than 20 papers on top-tier conferences and journals, such as INFOCOM, VLDB, IJCAI, IEEE Transactions on Parallel and Distributed Systems, and IEEE Transactions on Knowledge and Data Engineering.
\end{IEEEbiographynophoto}

\begin{IEEEbiographynophoto}{Yun~Peng}
is a full professor with the Institute of Artificial Intelligence and Blockchain, Guangzhou University. He received the PhD degree from the Hong Kong Baptist University, Hong Kong, China. His research interests are big data management, machine learning, data privacy, etc. He has published more than 20 papers on top-tier conferences and journals, such as SIGMOD, PVLDB, ICDE, IJCAI, and TKDE.
\end{IEEEbiographynophoto}

\begin{IEEEbiographynophoto}{Weilong~Peng}
received the Ph.D. degree from Tianjin University, Tianjin, China. He is currently with the School of Computer Science and Network Engineering, Guangzhou University, Guangzhou, China. His current research interests include computer vision and machine
learning.
\end{IEEEbiographynophoto}

\begin{IEEEbiographynophoto}{Kechao~Cai}
is an assistant professor at Sun Yat-Sen University. From the October 2019 to the March 2021, he was a postdoctoral research fellow at The Chinese University of Hong Kong. He received the Ph.D. degree in Computer Science supervised by Prof. John C.S. Lui at The Chinese University of Hong Kong in 2019. His research interests include data mining, data analytics, reinforcement learning, etc.
\end{IEEEbiographynophoto}

\end{document}